%% file: nc0.tex
\documentclass[12pt,a4paper]{article}
\usepackage{latexsym}
\usepackage{slashed}
\newcommand{\imu}{{\rm i}}
\newcommand{\md}{{\rm d}}
\newcommand{\Tr}{\mathop{\rm Tr}\nolimits}
\newcommand{\ster}[1]{\stackrel{\displaystyle #1}{,}}
 
\makeatletter 
\@addtoreset{equation}{section} 
\makeatother 
 
\begin{document} 
 
\title{\Huge\bf Non-Commutative Extensions \\ 
of the Standard Model\\[1cm]}  
\author{{\it Sorin Marculescu} \footnote{\tt marculescu\char64
physik.uni-siegen.de}  \\ 
Fachbereich Physik, Universit\"{a}t Siegen,
D-57068 Siegen, Germany} \date{} 
\maketitle 
 
\vspace{3cm} 
 
\begin{abstract} 
Four different extensions of the Standard Model to non-commutative 
space-time are considered. They all have the structure group 
~$U_{\rm Y}
( 1 )\bigotimes SU_{\rm L} ( 2 )\bigotimes SU_{\rm c} ( 3 )$~ but 
differ 
through the way Yukawa interaction is implemented. Models based on
non-commutative tensor products involve, in general, several inequivalent 
Seiberg--Witten maps of some (Higgs or fermionic) matter field. The 
non-minimal Non-Commutative Standard Model, advocated by the Munich Group, is
reproduced at lowest order in the non-commutativity parameter by a particular
model of this class.  
On the other hand, models based on hybrid Seiberg--Witten maps predict
electromagnetic couplings of neutral particles like $Z$-boson, Higgs meson, or
neutrino. The non-commutative contributions of the above Standard Model
extensions at low energies are evaluated by integrating out 
all massive bosonic degrees of freedom.    
\end{abstract} 
\newpage
\input{nc1}
%\newpage
\input{nc2}

%\newpage
\input{nc3}

%\newpage
\input{nc4}

%\newpage
\input{nc5}
%\newpage
\input{nc6}
%\newpage
\input{ncc}

\newpage
\appendix
\input{nc7}

\end{document}

%% file: nc1.tex
\section{Introduction}

In non-commutative field theories \cite{rev}
the space-time coordinates 
$x^{\mu}$ are promoted to Hermitean operators ${\hat x}^{\mu}$ 
obeying the commutations relations
\begin{equation}
\left[ {\hat x}^{\mu} , {\hat x}^{\nu} \right] = \imu 
{\theta}^{\mu \nu} \; \label{a1}
\end{equation}
where ${\theta}^{\mu \nu}$ is a real, antisymmetric matrix that may 
be a constant, a function of the coordinate operators, or a function 
of both coordinate and momentum operators.
While such a description has been proposed already in the early days 
of quantum field theory \cite{Sn} its present revival is mainly 
due to the 
development of nonperturbative aspects of string theory \cite{CDS}.

Non-commutative field theories are conveniently studied by replacing the
original multiplication law pointwise  by a deformed one, that reduces to the 
commutative multiplication in the limit of vanishing ${\theta}^{\mu \nu}$.
In the following a constant ${\theta}^{\mu \nu}$ will be 
considered, i.e. $\left[ {\hat x}^{\rho}, {\theta}^{\mu \nu} 
\right] = 0$. Accordingly, one introduces an non-commutative derivative 
 ${\hat \partial}_{\mu}$ with the properties
\begin{equation}
\left[ {\hat \partial}_{\mu} , {\hat \partial}_{\nu} \right] = 0 
\; , \qquad \qquad \left[ {\hat \partial}_{\mu} , {\hat x}^{\nu} \right] = 
{\delta}^{\nu}_{\mu} \; . \label{a6}
\end{equation}
It is convenient to replace ${\hat x}^{\mu}$ and 
${\hat \partial}_{\mu}$ by commuting coordinates $x^{\mu}$ and derivatives 
${\partial}_{\mu}$ defined by \cite{C}
\begin{equation}
x^{\mu} \equiv  {\hat x}^{\mu} - \frac{\imu}{2} {\theta}^{\mu \nu} 
{\hat \partial}_{\nu} \; , \qquad \qquad \; {\partial}_{\mu} \equiv 
{\hat \partial}_{\mu} \; .
\end{equation}  
A function ${\hat F} ( {\hat x} )$ of non-commutative coordinates becomes a
differential operator acting upon commuting coordinates  
\begin{equation}
{\hat F} ( {\hat x} ) = {\hat F} ( x ) \exp \left( 
\frac{\imu}{2} \stackrel{\leftarrow}{\partial}_{\mu} 
{\theta}^{\mu \nu} \stackrel{\rightarrow}{\partial}_{\nu} \right)
\equiv {\hat F} ( x ) \star  \; . \label{a2}
\end{equation}
The operator on the right of ${\hat F}$ is known as the Moyal 
(star) product operator \cite{M}.
It is easy now to check that the commutation relations (\ref{a1}) and
(\ref{a6}) are fulfilled.
The Moyal product considerably simplifies the quantization 
problem and the derivation of Feynman rules for NC field theories.
 
If the quantization takes into account the full 
nonlinear dependence on ${\theta}^{\mu \nu}$,
non-com\-mu\-ta\-tive field theories
show unusual properties,  like ultraviolet-infrared (UV / IR) mixing
\cite{MvRS},  or problems with unitarity for time-like \cite{{GM},{ABZ}} and
even space-like \cite{GGR} non-commutativity parameters.
By expanding the non-commutative field theory in ${\theta}^{\mu \nu}$, one 
obtains at each 
order a local field theory. If the asymptotic fields belong to the 
commutative region 
of space-time, the result is a unitary theory, free of UV /IR 
mixing, but non-renormalizable since   
${\theta}^{\mu \nu}$ is dimensionful. 

Representing non-commutative fields by Seiberg--Witten maps \cite{SW}, a 
method of implementing any gauge theory with compact structure group in
non-commutative space-time has been developed in refs. \cite{JSSW} and 
\cite{JMSSW}. Remarkably enough, the resulting non-commutative gauge theory
has the same one-loop anomalies as their commutative counterparts 
\cite{BMR}. When applied to the full Standard Model of particle physics the 
method leads to a minimal non-commutative extension \cite{Cal} with the same
number of coupling constants as the original Standard Model. 

However this extension is not unique. While no additional fields have to be
incorporated, there remains ambiguities in the choice of kinetic terms for the
gauge potentials and for the matter fields, including Higgs bosons. For
instance, a non-minimal Non-Commutative Standard Model (nmNCSM) has been
proposed in  \cite{Cal} and \cite{AJSW}, the main difference to the minimal
extension being due to the freedom of choosing the traces in the gauge kinetic
term. The phenomenological implications of both minimal and non-minimal
extensions have been considered in a number of works
\cite{{RS-J},{HPR},{BDDSTW},{CDR},{T},{OR},{XC},{MPT}}. 

A systematic discussion of the choice for matter kinetic terms is to our
knowledge still missing (see however \cite{{AJSW},{ST}}). In this paper
we propose to use in the construction of the matter kinetic terms only those
Seiberg--Witten maps required by the non-commutative Yukawa terms.  

To start with, we review in Sect. 2 the construction via 
Seiberg--Witten (SW) maps \cite{SW}
of the simplest non-commutative gauge theory. It consists of fermionic 
matter field minimally coupled to a gauge field. Beside gauge 
invariance the theory possesses a scaling symmetry that allows for 
simplifying the form of the action. This
scaling symmetry turns out to be useful in checking the consistency of  
the various Standard Model extensions. 
If the vector bosons acquire a finite mass, one can integrate out these fields 
and obtain the low energy behaviour of the model. We conclude this
section by computing the non-commutative contribution to the  low-energy
effective Hamiltonian. It describes four- and six-fermion processes.

In Sect. 3 we review the implementation of the Yukawa interaction  
by hybrid Seiberg--Witten maps \cite{Cal} and present an alternative 
realization through the non-commutative tensor product \cite{AJSW}. Some issues
concerning  the choice of Seiberg--Witten maps for matter fields (fermionic and
Higgs) are also  discussed.

In Sect. 4 we construct two different extensions of the Standard Model, in
which left-handed chiral fermions are represented by non-commutative tensor
fields.  In one model inequivalent Seiberg--Witten maps represent the Higgs
field $\varphi$ and its charge conjugate $\widetilde{\varphi}$. For both maps
we assume weighted contributions to the extended action. We find that, at
first  order in ${\theta}^{\mu \nu}$, the nmNCSM is a special case. 
In a second model the Seiberg--Witten maps
representing $\varphi$  and $\widetilde{\varphi}$ are related by complex
conjugation. As a result to the same left-handed chiral quark field one must
associate a pair of non-commuting tensor fields, complex conjugate each other, 
having weighted contributions to the kinetic term.
  
For comparison, we present in Sect. 5 two models based upon hybrid 
Seiberg--Witten maps. However, in contrast to ref.\cite{Cal},  we use such
maps through the whole non-com\-mu\-ta\-tive
extension of the Standard Model action 
and not only in Yukawa couplings. For this reason, the first model of this
section is somewhat different from the nmNCSM of refs.  \cite{Cal} and
\cite{MPTSW}. In particular, it predicts an electromagnetic interaction of
$Z$-bosons \cite{CPST} and of Higgs mesons \cite{GY}. In the second model the
left-handed chiral fermions are represented by hybrid Seiberg--Witten maps.
Such a construction has been used previously in GUT inspired models 
\cite{AJSW} and it is known to predict a coupling of the photon to neutrinos
\cite{{CPST},{STWR},{MST}} with interesting  astrophysical implications.
Since a hybrid map does not possess its own gauge field, its contribution to
the gauge sector may seem questionable. Based upon this observation we
propose to construct the kinetic gauge term in strict accordance with the
kinetic matter sector and to exclude contributions from all non-commuting gauge
fields that are not associated with a specific Seiberg--Witten map for a
matter multiplet. 

In Sect. 6 we integrate out all massive bosonic degrees of freedom of the 
Standard Model extensions in order to obtain their low-energy behaviour. We 
express in terms of physical fields the dominant non-commutative contribution
to the effective interaction Hamiltonians for the models presented above.

Our conclusions are reported in the last section.  
Some considerations about effective actions are collected in the Appendix. 

All models presented in the paper share some common features. On one hand,
they predict processes which are absent in the Standard Model. On the other
hand, they provide new contributions to Standard Model processes. In
principle, using various experimental bounds and precision experiments, one
should be able to distinguish between different non-commutative models,
although in practice such a test is still difficult.

Like in the Standard Model there are no gauge anomalies in our models and the
couplings to the non-commutative gravity are also anomaly free, at least in
versions of non-commutative gravity based upon Seiberg--Witten maps. 

The Higgs mechanism is not affected by the ${\theta}^{\mu \nu}$ expansion and
the custodial $SU(2)$ symmetry plays the same role as in the commutative
Standard Model. In particular, it allows for parameterizing the low-energy
effective Hamiltonian in terms of the Fermi coupling constant.

We conclude this Introduction by the following remarks: While the constructions
of the Seiberg--Witten  maps can be in principle
performed to any finite order in ${\theta}^{\mu \nu}$, we will restrict the
explicit computations to first order. 
We shall deal exclusively with the (old) Standard Model in which all neutrinos
are massless. There is however, in principle, no problem to incorporate massive
neutrinos. 

%% file: nc2.tex
\section{Review of Seiberg--Witten Maps}

The Moyal product of two functions is a power series in the 
non-commutativity parameter ${\theta}^{\mu \nu}$ starting with the 
commutative product plus higher order terms chosen in such a way as 
to yield an associative product. It can be used to express 
infinitesimal non-commutative gauge transformations of matter and 
gauge fields
\begin{equation}
\delta {\hat \Psi} = \imu {\hat \Lambda} \star {\hat \Psi} \; , \qquad
\qquad \delta {\hat V}_{\rho} = {\partial}_{\rho} {\hat \Lambda} + 
\imu \left[ 
{\hat \Lambda} \ster{\star} {\hat V}_{\rho} \right] \; \label{b1}
\end{equation}
under usual gauge transformations
\begin{equation}
\delta \psi = \imu \lambda \psi \; , \qquad \qquad  \delta v_{\rho} = 
{\partial}_{\rho} \lambda - \imu \left[ v_{\rho} , \lambda \right] 
\equiv D_{\rho} \lambda \; . 
\end{equation}
Furthermore, any pair of non-commutative gauge transformations ${\hat
\Lambda}$,  ${\hat \Xi}$ has to satisfy the consistency condition \cite{JMSSW}
\begin{equation}
\imu \left( {\delta}_{\lambda} {\hat \Xi} - {\delta}_{\xi} 
{\hat \Lambda} \right) + \left[ {\hat \Lambda} \ster{\star} {\hat \Xi} 
\right] = \widehat{\left[ \lambda , \xi \right]} \; . \label{b2}
\end{equation}
In (\ref{b1}), (\ref{b2}) the symbol $\star$ includes 
also matrix multiplication. 
The suffixes $\lambda$, $\xi$ mean that variations are performed 
with respect to two different gauge transformation and  
the right hand side of (\ref{b2}) denotes a non-commutative gauge
transformation  of the commutator $\left[ \lambda, \xi \right]$.
 
Commutative gauge fields and parameters are Lie algebra valued. 
The matter field $\psi$ belongs to an arbitrary (unitary) 
representation of a compact gauge group with generators $t_a$ 
satisfying 
\begin{equation}
\left[ t_a , t_b \right] = \imu C^{a b c} t_c \; . 
\end{equation}
Since the 
Moyal commutator, as introduced in (\ref{b1}) and (\ref{b3}) includes the
anticommutator $\left\{ t_a , t_b  \right\}$, non-commutative fields and
parameters do not belong in general to the Lie algebra, they are  in the
enveloping of the Lie algebra.

The Seiberg--Witten maps of the gauge theory are local solutions of
(\ref{b1}),  (\ref{b2}). This means that they can be expressed as
formal series in ${\theta}^{\mu \nu}$ that at 
each order depend on commutative fields and a finite number of 
their derivatives.
To first order in ${\theta}^{\mu \nu}$ one finds
\begin{eqnarray}
{\hat \Lambda} & = & \lambda + \frac{1}{2} {\theta}^{\mu \nu} 
\left( c {\partial}_{\mu} \lambda v_{\nu} + c^{\ast} v_{\nu} 
{\partial}_{\mu} \lambda \right) \;  , \\
{\hat \Psi}     & = & \psi + \frac{1}{2} {\theta}^{\mu \nu} 
\left[  v_{\mu} \left( - {\partial}_{\nu} + \imu c v_{\nu} \right) 
+ a v_{\mu \nu} \right] \psi \; , \label{b3}\\
{\hat V}_{\rho} & = & v_{\rho} + \frac{1}{2} {\theta}^{\mu \nu} 
\left\{ - \frac{1}{2} \left\{ v_{\mu} , {\partial}_{\nu} v_{\rho} +
v_{\nu \rho} \right\} + D_{\rho} \left[ b v_{\mu \nu} + 
\left( c - \frac{1}{2} \right) v_{\mu} v_{\nu} \right] \right\} \;
,\label{b4}  \end{eqnarray} 
where $a$ can be complex and $b$ is real. The parameter 
$c$ is gauge dependent and will be chosen $1 / 2$. 
In contrast to $c$, the parameters $a$ and $b$ 
multiply covariant quantities and may be present in the invariant 
action. Since they can be eliminated by a (covariant) field 
redefinition we call them scaling parameters.  

From (\ref{b3}), (\ref{b4}) one can construct Seiberg--Witten maps for
covariant  derivatives and  field strengths
\begin{equation}
{\hat {\cal D}}_{\rho} \star {\hat \Psi} 
 = {\partial}_{\rho} {\hat \Psi} -  \imu {\hat V}_{\rho} \star 
{\hat \Psi} \; , \quad \qquad \;
{\hat V}_{\rho \sigma} 
 = {\partial}_{\rho} {\hat V}_{\sigma} - {\partial}_{\sigma} 
{\hat V}_{\rho} - \imu \left[ 
{\hat V}_{\rho} \ster{\star} {\hat V}_{\sigma} \right] \;     
\end{equation}
whose infinitesimal non-commutative transformations are
\begin{equation}
\delta {\hat {\cal D}}_{\rho} \star {\hat \Psi} 
 =  \imu {\hat \Lambda} \star \left( {\hat {\cal D}}_{\rho} \star 
{\hat \Psi} \right) \; , \quad \qquad \;
\delta {\hat V}_{\rho \sigma} 
 =  \imu \left[ 
{\hat \Lambda} \ster{\star} {\hat V}_{\rho \sigma} \right] \; .    
\end{equation}

By assuming usual boundary conditions at infinity and by using Stokes 
theorem one can immediately show that the action integral
\begin{equation}
{\cal S} = \int {\md}^4 x \left[ {\bar{\hat \Psi}} \star \imu 
\left( {\hat {\slashed{\cal D}}} \star 
{\hat \Psi} \right) - M {\bar{\hat \Psi}} \star {\hat \Psi}  - 
\frac{1}{2 g^2} \Tr {\hat V}_{\rho \sigma} \star 
{\hat V}^{\rho \sigma} \right] \; \label{b5}  
\end{equation}
is gauge invariant. 

An evaluation of (\ref{b5}) up to first order in 
${\theta}^{\mu \nu}$ leads to
\begin{equation}
{\cal S} = {\cal S}^0 + \Delta {\cal S}_g + \Delta {\cal S}_m \;
\end{equation}
where ${\cal S}^0$ is the classical commutative action and   
$\Delta {\cal S}_g$, $\Delta {\cal S}_m$ are the non-commutative contributions 
to the gauge sector and to the matter sector, respectively. They are 
given by
\begin{eqnarray}
{\cal S}^0        & = & {\cal S}^0_m + {\cal S}^0_g = \int {\md}^4 x 
\left[ \bar{\psi} \left( \imu \slashed{D} - M \right) \psi  
- \frac{1}{2 g^2} \Tr v_{\rho \sigma}v^{\rho \sigma} \right] \; , 
 \\ 
\Delta {\cal S}_g & = & \frac{1}{g^2} \int {\md}^4 x  \frac{1}{2} 
{\theta}^{\mu \nu} 
\Tr v^{\rho \sigma} \left( - 2 v_{\mu \rho} v_{\nu \sigma} + 
\frac{1}{2} v_{\rho \sigma} v_{\mu \nu} \right) \; , \label{b6} \\
\Delta {\cal S}_m & = & \int {\md}^4 x \frac{1}{2} {\theta}^{\mu \nu}
\left[ \bar{\psi} \imu 
{\gamma}^{\rho} v_{\mu \rho} D_{\nu} \psi + \left( a - \imu b \right) 
\left( - \imu D_{\rho} 
\bar{\psi} {\gamma}^{\rho} - M \bar{\psi} \right) v_{\mu \nu} \psi
\right. \nonumber \\
                  &    & \mbox{} + \left.\left( a^{\ast} + \imu b 
- \frac{1}{2} \right) \bar{\psi} v_{\mu \nu} \left( \imu \slashed{D}
- M \right) \psi \right] \; . \label{b7}
\end{eqnarray}    
Notice the presence of the 
parameters $a$ and $b$ only in the matter action where they multiply 
the equation of motion and can be scaled away by the following field 
redefinitions:
\begin{equation}
\psi \longrightarrow \psi - \frac{a}{2} {\theta}^{\mu \nu} v_{\mu \nu} \psi 
\; , \quad \qquad \; v_{\rho} \longrightarrow v_{\rho} - \frac{b}{2} 
{\theta}^{\mu \nu} D_{\rho} v_{\mu \nu} \; . 
\end{equation} 
The resulting non-commutative contribution is
\begin{equation}
\Delta {\cal S}_m = - \int {\md}^4 x \left( \frac{\imu}{2} \bar{\psi}  
{\theta}^{\mu \nu \rho} v_{\mu \nu} D_{\rho} \psi - \frac{1}{2} 
{\theta}^{\mu \nu} \bar{\psi} M v_{\mu \nu} \psi \right) \; 
\end{equation}
where we introduced the completely antisymmetric tensor
\begin{equation}
{\theta}^{\mu \nu \rho} \equiv \frac{1}{2} \left( {\theta}^{\mu \nu} 
{\gamma}^{\rho} + {\theta}^{\nu \rho}{\gamma}^{\mu} + {\theta}^{\rho \mu}
{\gamma}^{\nu} \right) \; .
\end{equation}
The independence on scaling parameters has been verified  up to 
second order in ${\theta}^{\mu \nu}$, in ref. \cite{Moe}.

Nontrivial physics is encoded in the (matter) currents. They are defined 
through the variation of the matter action with respect to the gauge 
fields.
For the gauge field theory under consideration we have
\begin{equation}
- {\delta}_v {\cal S}^0_m \equiv \int {\md}^4 x J_a^{\rho} \delta 
A_{\rho}^a \; , \qquad v_{\rho} = - g A_{\rho}^a t_a \; ,\qquad
J_a^{\rho} = g \bar{\psi} {\gamma}^{\rho} t_a \psi \; 
\end{equation}  
where we introduced explicitly the gauge coupling constant $g$.
The non-commutative contribution to the matter current can be obtained from 
\begin{equation}
{\delta}_v \Delta {\cal S}_m \equiv \int {\md}^4 x  \Delta J_a^{\rho} ( v ; x
)  \delta A_{\rho}^a ( x )\; \label{b8}
\end{equation}
where  
\begin{equation}
\Delta J_a^{\rho} ( v ) = g \left[ \bar{\psi} {\theta}^{\mu \nu \rho}
\left( \imu \stackrel{\leftarrow}D_{\mu} t_a D_{\nu} + \frac{1}{2} 
\left\{ t_a , v_{\mu \nu} 
\right\} \right) + {\theta}^{\mu \rho} D_{\mu} \left( \bar{\psi} M t_a \psi 
\right) \right] \; . \label{b9}
\end{equation}
The notation emphasizes the additional dependence upon the  gauge field.
 
If some of the gauge fields acquire mass the low energy
behavior is adequately described by an effective theory obtained by
integrating out  the massive gauge fields. Since the gauge fields couple
linearly to fermions the fist term in the effective action has the 
current-current  form 
\begin{equation}
{\cal H}_0^{\rm eff} ( x ) = \frac{1}{2} J_a^{\mu} ( x ) {\bar J}_{\mu a} ( x )
+ \cdots \; .  \label{b11}
\end{equation}
where
\begin{equation}
{\bar J}_{\mu a} ( x ) \equiv \int {\md}^4 y D_{\mu \nu} ( x - y ) J_a^{\nu} (
y ) \; ,
\end{equation}
with $D_{\mu \nu} ( x )$ the free propagator of the massive gauge field
\begin{equation}
D_{\mu \nu} ( x ) \equiv \int \frac{{\md}^4 k}{( 2 \pi )^4} 
{\rm e}^{- \imu k x}  
\displaystyle{\frac{{\eta}_{\mu \nu} - \frac{k_{\mu} k_{\nu}}{m^2}}{m^2 -
k^2}}  \; , \label{b12} 
\end{equation}
and the dots stay for terms of at least cubic order in ${\bar J}_{\mu a}$. 

To get a local effective
theory one has to expand the propagator in inverse  powers of the gauge field
mass: 
\begin{equation}
{\bar J}_{\mu a} ( x ) = \frac{1}{m^2} J_{\mu a} ( x ) + \frac{1}{m^4} \left( 
{\partial}_{\mu} {\partial}_{\nu} - {\eta}_{\mu \nu} \Box \right) \left( 1 - 
\frac{\Box}{m^2} + \cdots \right) J_a^{\nu} ( x ) \; . 
\end{equation}
The effective action itself is obtained as an expansion in $1 / m^2$.
Actually, it makes sense to keep in this expansion only terms which do not
require further regularization, such that the effective action will be
expressed solely through bare parameters of the original theory.  

Since the non-commutative current has additional gauge field dependence the 
evaluation of the low energy effective action is more involved. We assume of
course, that the occurrence of a vector boson mass term in the commutative
sector is the only modification suffered by (\ref{b5}). Due to the 
antisymmetric tensor ${\theta}^{\mu \nu \rho}$, the contribution induced 
by $\Delta {\cal S}_m$  remains unrenormalized and is given by
\begin{eqnarray} \Delta {\cal H}_{\rm NC}^m    
& = & g \left[ \imu
{\partial}_{\mu} \bar{\psi}  {\theta}^{\mu \nu \rho} t_a {\partial}_{\nu} \psi
+ {\theta}^{\mu \rho}  {\partial}_{\mu} \left( \bar{\psi} M t_a \psi \right)
\right] {\bar J}_{\rho a} \;  \nonumber \\ 
&   & - \mbox{}  \frac{g^2}{2}
{\bar J}_{\rho a} \left\{ C^{a b c} \bar{\psi}   \left[ \frac{\imu}{2}
{\theta}^{\mu \nu \rho} \left( \stackrel{\leftarrow}{\partial}_{\mu} -
\stackrel{\rightarrow} {\partial}_{\mu} \right)  +  {\theta}^{\nu \rho} M
\right] t_c \psi +  \bar{\psi} {\theta}^{\mu \nu \rho} \left\{ t_a , t_b
\right\} \psi  \partial_{\mu}  \right\} {\bar J}_{\nu b} \; \nonumber \\
&   & - \mbox{} \imu g^3  \bar{\psi} {\theta}^{\mu \nu \rho}  t_a t_b t_c
\psi {\bar J}_{\mu b}{\bar J}_{\nu c}{\bar J}_{\rho a} \; . \label{b14}
\end{eqnarray}
Currents being bilinear in the fermionic fields, (\ref{b14}) describe 
four-, six- and eight-fermion effective interactions. 

On the other hand, $\Delta {\cal S}_g$ contains triple and quadruple gauge
field interactions, which also describe processes with more than four
fermions. As shown in the Appendix, quartic gauge field interactions provide
contributions of order $1 / m^6$ to effective four-fermion processes
depending on the regularization details. The non-commutative contribution to
the effective Hamiltonian becomes thus
\begin{eqnarray}
\Delta {\cal H}_{\rm NC}^{\rm eff} 
& = &  \frac{g^2}{m^2} \left[ \imu {\partial}_{\mu}\bar{\psi} 
{\theta}^{\mu \nu \rho} t_a {\partial}_{\nu} \psi + {\theta}^{\mu \rho} 
{\partial}_{\mu} \left( \bar{\psi} M t_a \psi \right) \right] \left[ 
\eta_{\rho \sigma} + \frac{1}{m^2} \left( 
{\partial}_{\rho} {\partial}_{\sigma} - {\eta}_{\rho \sigma} \Box \right)
\right]{\bar \psi} {\gamma}^{\sigma} t_a \psi \; \nonumber \\
&   & \mbox{} - \frac{g^4}{2 m^4} {\bar \psi} {\gamma}_{\rho} t_a \psi   
\left\{ C^{a b c} \bar{\psi} 
\left[ \frac{\imu}{2} {\theta}^{\mu \nu \rho}
\left( \stackrel{\leftarrow}{\partial}_{\mu} - \stackrel{\rightarrow}
{\partial}_{\mu} \right)  +  {\theta}^{\nu \rho} M \right] t_c \psi 
\right. \; \nonumber \\ 
&   & \mbox{} + \left. \bar{\psi} {\theta}^{\mu \nu \rho} \left\{ t_a , t_b
\right\} \psi  \partial_{\mu}  \right\} {\bar \psi} {\gamma}_{\nu} t_b \psi 
+ {\cal O} \left( \frac{1}{m^6}   \right) \; . \label{b13}
\end{eqnarray}
Notice that, in contrast with the commutative case the part of (\ref{b13})
describing six-fermion interactions  is of order $1 / m^4$ and does not
depend upon the regularization details.

%% file: nc3.tex
\section{Yukawa Interaction and Seiberg--Witten Maps}

The action of the commutative Standard Model consists of several parts
representing the pure gauge ($g$) sector and the matter sectors (two fermionic
sectors labelled by the superscript $m$ and one Higgs sector labelled by $H$):
 \begin{equation}
{\cal S}_{\rm SM} = {\cal S}^{(g)} + {\cal S}_{\rm matter} = {\cal S}^{(g)} + 
\left[ \sum_m {\cal S}^{(m)} + {\cal S}^{(H)} \right] \; .
\end{equation}
It is convenient to write the contribution of the matter sectors in the form 
\begin{eqnarray}
{\cal S}_{\rm matter} & = & \int {\md}^4 x \left\{ \sum_m \left[
{\bar \psi}_{\rm L} \imu \slashed{D} \psi_{\rm L} + {\bar \chi}_{\rm R} \imu 
\slashed{D} \chi_{\rm R} +  {\bar \eta}_{\rm R} \imu \slashed{D} 
\eta_{\rm R} - \left( {\bar \psi}_{\rm L} G^{+} \varphi \chi_{\rm R}
\right. \right. \right.\; \nonumber \\  
&& \mbox{}  + \left. \left. \left.
{\bar \psi}_{\rm L} {\widetilde G}^{+} \widetilde{\varphi} 
\eta_{\rm R} + {\rm h}.{\rm c}. \right) \right]
 + D^{\rho} \varphi^{+} D_{\rho} \varphi + 
\frac{\lambda v^2}{4} {\varphi}^{+} \varphi - 
\frac{\lambda}{4} ({\varphi}^{+} \varphi )^2 
 \right\} \; \label{c0}
\end{eqnarray} 
where we omit the label $m$ assigned to each fermionic field.
Spinor and flavor indices are not shown explicitly, summation being
understood. The fermions $\chi$ and $\eta$ are decomposed  into their left
(suffix ${\rm L}$) and right (suffix ${\rm R}$) chiral parts, the Higgs field
is denoted  by $\varphi$, its charge conjugated by $\widetilde{\varphi} \equiv 
\imu \tau^2 \varphi$. ${\psi}_{\rm L}$ and $\varphi$ form the doublets
\begin{equation}
 \psi_{\rm L} = \left[ \begin{array}{c}
                         \eta_{\rm L} \\ \chi_{\rm L}
                          \end{array} 
                   \right] \; , \qquad \qquad  \; 
    \varphi  = \frac{1}{\sqrt{2}}
                   \left[ \begin{array}{c}
                          0   \\ v + H 
                          \end{array}
                    \right] \; 
\end{equation}
with $H$ describing the physical Higgs meson. 
The gauge
field associated to the matter multiplet labelled by $n$ is
\begin{equation}
- v_{\rho} ( n ) \equiv g' Y_n B_{\rho} + g T_n^a W_{\rho}^a
+ g_s T_n^A G_{\rho}^A \; 
\end {equation}
where $Y_n$, $T_n^a$, $T_n^A$ denote weak hypercharge, weak isospin
and $SU(3)$ color, respectively. We shall use occasionally the abbreviations
\begin{equation}
v_{\rho} = v_{\rho} ( \varphi ) \; , \quad \qquad
v_{{\rm R} \rho} = v_{\rho} ( {\chi}_{\rm R} ) \; , \quad \qquad
v_{{\rm R} \rho}^{'} = v_{\rho} ( {\eta}_{\rm R} ) \; .
\end{equation}
 The 
gauge field $v_{{\rm L} \rho}$ associated to the left chiral 
doublet ${\psi}_{\rm L}$ can be computed from
\begin{equation}
v_{{\rm L} \rho} = v_{\rho} + v_{{\rm R} \rho} = 
\widetilde{v}_{\rho} + v_{{\rm R} \rho}^{'} \; \label{c1} 
\end{equation}
where ${\widetilde{v}}_{\rho} \equiv - \tau^2 v_{\rho}^{\ast} \tau^2$ 
is the gauge field associated to the charge conjugated Higgs.

The gauge fields $B_{\rho}$, $W_{\rho}^a$ are expressed through physical ones
by \begin{equation}
Z_{\rho} = \cos \theta W_{\rho}^3 - \sin \theta B_{\rho} \; , \qquad  
A_{\rho} = \sin \theta W_{\rho}^3 + \cos \theta B_{\rho} \; , \qquad  
W_{\rho}^{\pm} = \frac{W_{\rho}^1 \mp \imu W_{\rho}^2}{\sqrt{2}} \;
\end{equation}
with $\theta$, the Weinberg angle.
 
The physical matter fields $\chi$ and $\eta$ are obtained with help of unitary
matrices $A_{\rm L , \rm R}$, $A_{\rm L , \rm R}^{'}$ and chiral projection 
operators $P_{\rm L , \rm R} \equiv 1 / 2 \left( 1 \mp \gamma_5 \right)$:
\begin{equation}
{\chi}_{\rm L , \rm R} = A_{\rm L , \rm R} P_{\rm L , \rm R} \chi \; , \qquad
\qquad {\eta}_{\rm L , \rm R} = A_{\rm L , \rm R}^{'} P_{\rm L , \rm R} \eta 
\; ,
\end{equation}
such as to have diagonal mass matrices
\begin{equation}
M \equiv A_{\rm R}^{+} \frac{G v}{\sqrt{2}} A_{\rm L} \; , \qquad \qquad 
M' \equiv A_{\rm R}^{' +} \frac{\widetilde{G} v}{\sqrt{2}} A_{\rm L}^{'} \; .
\end{equation}
When both fields are massive a Cabibbo--Kobayashi--Maskawa matrix 
$V \equiv  A_{\rm L}^{' +} A_{\rm L}$ can be introduced.

We will write the action for the pure Yang--Mills part of 
the Standard Model in the form
\begin{equation}
{\cal S}^{(g)} = - \frac{1}{2} \int {\md}^4 x \sum_n \frac{1}{h_n^2} \Tr 
v_{\rho \sigma} ( n ) v^{\rho \sigma} ( n ) \; 
\end{equation}
where certain weights, $1 / h_n^2$, are assigned to the contribution of each 
of the gauge fields in the matter representation of the gauge group labelled
by $n$. As we shall see in the next sections, what matter representations enter
the sum may depend upon the specific non-commutative extension.  The
normalization conditions are 
\begin{equation}
\frac{1}{2 g^{' 2}} = \sum_n \frac{Y_n^2}{h_n^2} \; , \qquad 
\frac{1}{2 g^2} = \sum_n \frac{\Tr \left( T_n^a T_n^a \right)}{h_n^2} \; , 
\qquad  
\frac{1}{2 g_s^2} = \sum_n \frac{\Tr \left( T_n^A T_n^A \right)}{h_n^2} 
\; . \label{c17}    
\end{equation}

While most of the terms in (\ref{c0}) have a straightforward
non-commutative extension, for  the Yukawa interaction
\begin{equation}
- \bar{\psi}_{\rm L} G^{+} \varphi {\chi}_{\rm R} + {\rm c.c.} \; , 
\label{c2}
\end{equation}
there are several independent possibilities.
The authors of ref. \cite{Cal} represent the non-com\-mu\-ta\-tive Higgs field 
${\hat \Phi}$ by a hybrid Seiberg--Witten map transforming left and right,  
under left- and right-handed chiral gauge groups, respectively
\begin{equation}
\delta {\hat \Phi} = \imu {\hat \Lambda}_{\rm L} \star {\hat \Phi} 
- \imu {\hat \Phi} \star {\hat \Lambda}_{\rm R} \; . \label{c3}
\end{equation}
By assuming the non-commutative gauge transformations  
\begin{equation}
\delta {\hat \Psi}_{\rm L} = \imu {\hat \Lambda}_{\rm L} \star 
{\hat \Psi}_{\rm L} \; ,
\qquad  \qquad \delta {\hat \chi}_{\rm R} = \imu {\hat \Lambda}_{\rm R} \star
{\hat \chi}_{\rm R} \;  \label{c14}
\end{equation}
for the maps ${\hat \Psi}_{\rm L}$ and ${\hat \chi}_{\rm R}$ associated to 
left-handed chiral and right-handed chiral fermions,  
one can check the gauge invariance of the following  extension of the
Yukawa interaction  (\ref{c2}):
\begin{equation}
{\cal S}_{\rm HSY} = - \int {\md}^4 x \left( \bar{\hat{\Psi}}_{\rm L} G^{+} 
\star {\hat \Phi} \star {\hat \chi}_{\rm R} + {\rm h.c.} \right) \; . 
\label{c4}
\end{equation}

The most general solution for the hybrid scalar Seiberg--Witten  
map up to first order in 
${\theta}^{\mu \nu}$ is given by 
\begin{eqnarray}
{\hat \Phi} & = & \phi + \frac{1}{2} {\theta}^{\mu \nu} 
\left[  v_{{\rm L} \mu} \left( - {\partial}_{\nu} + \imu c_{\rm L} 
v_{{\rm L} \nu} \right)\phi  +  \left( {\partial}_{\mu} \phi + \imu 
c^{\ast}_{\rm R} \phi v_{{\rm R} \mu} \right) v_{{\rm R} \nu} \right.
\; \nonumber \\
            &   & \mbox{} - \left.   
\imu v_{{\rm L} \mu} \phi v_{{\rm R} \nu} + \alpha_{\rm L} 
v_{{\rm L} \mu \nu} \phi
 + \alpha_{\rm R} \phi  v_{{\rm R} \mu \nu} \right]  \; \label{c16}
\end{eqnarray}
where $\alpha_{\rm L}$ and $\alpha_{\rm R}$ are arbitrary complex scaling
parameters. The gauge dependent constants $c_{\rm L}$ and $c_{\rm R}$ do not
show up in gauge invariant expressions and will be,as usual, chosen    
$1 / 2$. The contact with the commutative Higgs field 
$\varphi$  is done 
by representing the commutative part $\phi$ as the 
unit matrix ${\rm I}_3$ in color space $\phi = \varphi \bigotimes {\rm I}_3$.

An alternative to the use of hybrid Seiberg--Witten 
maps is the construction of non-com\-mu\-ta\-tive extensions of 
tensor fields  and tensor product of fields. The infinitesimal gauge
transformation, direct product of two commuting gauge transformations with
parameters  $\lambda$ and ${\lambda}_{\rm R}$, is represented in the
non-commutative space-time by the  map
\begin{eqnarray}
{\hat \Omega}_{( \lambda, {\lambda}_{\rm R} )} 
& = & \lambda + 
{\lambda}_{\rm R} + \frac{1}{2} {\theta}^{\mu \nu} \left[ c 
{\partial}_{\mu} \lambda v_{\nu} + c_{\rm R}{\partial}_{\mu} 
{\lambda}_{\rm R} v_{{\rm R} \nu} + {\rm h. c.} \right. \; \nonumber \\
&   & \mbox{} + \left. \left( 2 - 
{\gamma}_{\rm R} \right) {\partial}_{\mu} \lambda v_{{\rm R} \nu} 
+ {\gamma}_{\rm R}{\partial}_{\mu} {\lambda}_{\rm R} v_{\nu} \right] \; 
\label{c15}
\end{eqnarray} 
with ${\gamma}_{\rm R}$ a new gauge dependent real constant, independent on 
$c$ and $c_{\rm R}$. The gauge fields associated to the two commuting groups
are  denoted by $v_{\rho}$ and $v_{\rm R \rho}$.
This formula and subsequent considerations simplify considerably
in the gauge defined by $c = c_{\rm R} = 1 / 2$ and  ${\gamma}_{\rm R} = 1$.
We have from (\ref{c15})   
\begin{equation}
{\hat \Omega}_{( \lambda, {\lambda}_{\rm R} )} = \lambda + 
{\lambda}_{\rm R} + \frac{1}{4} {\theta}^{\mu \nu} \left\{  
{\partial}_{\mu} \left( \lambda + {\lambda}_{\rm R} \right) , v_{\nu}
+  v_{{\rm R} \nu} \right\}  \; .
\label{c5}
\end{equation} 
On the right-hand side of (\ref{c5}) one can recognize the Seiberg--Witten map 
${\hat \Lambda}_{\rm L}$ of left chiral gauge transformations 
${\lambda}_{\rm L} = {\lambda}_{\rm R} + \lambda$ with the corresponding 
gauge field $v_{{\rm L} \rho} = v_{\rho} + v_{{\rm R} \rho}$.

Both the non-commutative extension of tensor product 
$\varphi {\chi}_{\rm R}$  
\begin{eqnarray}
\widehat{\varphi {\chi}_{\rm R}} 
& = &  \varphi {\chi}_{\rm R} + \frac{1}{2} {\theta}^{\mu \nu} \left[ 
v_{\rm L \mu} \left( - {\partial}_{\nu} + \frac{\imu}{2} v_{\rm L \nu} \right)
\left( \varphi {\chi}_{\rm R} \right) + \imu D_{\mu}\varphi D_{\nu} 
{\chi}_{\rm R} \right. \; \nonumber \\
&   & \mbox{} + \left. \left( \alpha v_{\mu \nu} + {\alpha}_{\rm R} 
v_{\rm R \mu \nu}  \right)\varphi {\chi}_{\rm R} \right] \; \label{c9}
\end{eqnarray} 
and the Seiberg--Witten map of left-handed chiral fermion ${\psi}_{\rm L}$
\begin{equation}
\hat{\Psi}_{\rm L} 
 =  {\psi}_{\rm L} + \frac{1}{2} {\theta}^{\mu \nu} \left[ 
v_{\rm L \mu} \left( - {\partial}_{\nu} + \frac{\imu}{2} v_{\rm L \nu} \right)
+ \left( a_{\rm L} v_{\mu \nu} + a_{\rm L}^{'} v_{{\rm R} \mu \nu} 
\right) {\psi}_{\rm L} \right] \; \label{c10}
\end{equation} 
transform with the non-commutative parameter of the product gauge 
transformation (\ref{c5}) as
\begin{equation}
\delta \widehat{\varphi {\chi}_{\rm R}} = \imu 
{\hat \Omega}_{( \lambda, {\lambda}_{\rm R} )} \star 
\widehat{\varphi {\chi}_{\rm R}} \; , \quad \qquad
\delta \hat{\Psi}_{\rm L} = \imu 
{\hat \Omega}_{( \lambda, {\lambda}_{\rm R} )} \star 
\hat{\Psi}_{\rm L} \; ,
\end{equation}
under $\delta \varphi = \imu \lambda \varphi$ and $\delta 
{\chi}_{\rm R} = \imu {\lambda}_{\rm R} {\chi}_{\rm R} \;$.

The gauge invariant extension of the Yukawa interaction is in this case
\begin{equation}
{\cal S}_{\rm TPY} = - \int {\md}^4 x \left( \bar{\hat{\Psi}}_{\rm L} 
G^{+} \star \widehat{\varphi {\chi}_{\rm R}} + {\rm h.c.} \right) \; . 
\end{equation}

A further proliferation of non-commutative standard models is caused by the 
different couplings of chiral fermions to $\varphi$ and 
$\widetilde{\varphi}$. If the Seiberg-Witten maps of $\varphi$ and 
$\widetilde{\varphi}$ are independent of each other, the non-commutative
extension of the Yukawa coupling  involving $\widetilde{\varphi}$ is obtained
from the coupling of   $\varphi$ by replacing
the gauge fields $v_{\rho}$, $v_{\rm R \rho}$  and the gauge parameters
$\lambda$, $\lambda_{\rm R}$ through $\widetilde{v}_{\rho}$, $v_{\rm R
\rho}^{'}$ and  $\widetilde{\lambda}$, $\lambda_{\rm R}^{'}$ respectively. 
We shall argue in the next section that, due to (\ref{c1}), the 
non-commuting extensions of the direct product 
gauge transformations ${\hat \Omega}_{( \lambda, {\lambda}_{\rm R} )}$ and 
${\hat \Omega}_{( \widetilde{\lambda}, {\lambda}_{\rm R}^{'} )}$ can be 
identified. 

In case that Seiberg--Witten maps of the Higgs field and its charge conjugated
are related by complex conjugation, maps constructed with either the direct
Moyal product, or with the opposite product
\begin{equation}
\circ \equiv {( \star )}^{\ast} = \exp \left( - 
\displaystyle{\frac{\imu}{2}} \stackrel {\leftarrow}{\partial}_{\mu} 
{\theta}^{\mu \nu} \stackrel{\rightarrow}{\partial}_{\nu} \right) \; ,
\label{c18}
\end{equation} 
appear in the same non-commutative version of the Standard Model.

In the hybrid fermion model, to be discussed in Sect. 5 one starts with the
full Yukawa term in the form  
\begin{equation}
- {\bar \chi}_{\rm R} G {\psi}_{\rm L}^t {\varphi}^{\ast} -
{\bar \eta}_{\rm R} \widetilde{G} {\psi}_{\rm L}^t
{\widetilde{\varphi}}^{\ast} + {\rm c.c.} \;  \label{c12}
\end{equation}
where the superscript $t$ means transposition in the weak hypercharge-isospin
space. While the Seiberg--Witten map for ${\varphi}^{\ast}$ can be still
constructed with the usual star product by the choice 
$\widehat{{\Phi}^{\ast}} \equiv - \imu \tau^2 {\hat \Phi}_2$,  
with ${\hat \Phi}_2$ representing $\widetilde{\varphi}$ in 
non-commutative space-time,  the map of ${\widetilde{\varphi}}^{\ast}$  must
be complex conjugate to  ${\hat \Phi}_2$. As a consequence, 
Seiberg--Witten maps of  ${\chi}_{\rm R}$ and ${\eta}_{\rm R}$ have to be
constructed with opposite  products.  Also the left chiral fermion will be
represented by two maps ${\hat \Psi}_{1 \rm L}^t$ and 
${\hat \Psi}_{2 \rm L}^t$, opposite each other and hybrid of either 
${\varphi}^{\ast}$ and ${\chi}_{\rm R}$, or  ${\widetilde{\varphi}}^{\ast}$
and ${\eta}_{\rm R}$. The general solutions of the corresponding 
consistency conditions are
\begin{eqnarray}
{\hat \Psi}_{1 \rm L}^t 
& = & {\psi}_{\rm L}^t + \frac{1}{2} {\theta}^{\mu \nu} 
\left[  v_{{\rm R} \mu} \left( - {\partial}_{\nu} + \frac{\imu}{2}  
v_{{\rm R} \nu} \right){\psi}_{\rm L}^t  +  
\left( - {\partial}_{\mu} {\psi}_{\rm L}^t +
\frac{\imu}{2} {\psi}_{\rm L}^t v_{\mu}^{\ast} \right) 
v_{\nu}^{\ast} \right.
\; \nonumber \\
            &   & \mbox + \left.   
\imu v_{{\rm R} \mu} {\psi}_{\rm l}^t v_{\nu}^{\ast} + a_{1 \rm L}^{'} 
v_{{\rm R} \mu \nu} {\psi}_{\rm L}^t
 - a_{1 \rm L} {\psi}_{\rm L}^t  v_{\mu \nu}^{\ast} \right] \; , 
\label{c6} \\ 
{\hat \Psi}_{2 \rm L}^t 
& = & {\psi}_{\rm L}^t - \frac{1}{2} {\theta}^{\mu \nu} 
\left[  v_{{\rm R} \mu}^{'} \left( - {\partial}_{\nu} + \frac{\imu}{2}  
v_{{\rm R} \nu}^{'} \right){\psi}_{\rm L}^t  +  
\left( - {\partial}_{\mu} {\psi}_{\rm L}^t +
\frac{\imu}{2} {\psi}_{\rm L}^t \widetilde{v}_{\mu}^{\ast} \right) 
\widetilde{v}_{\nu}^{\ast} \right.
\; \nonumber \\
            &   & \mbox + \left.   
\imu v_{{\rm R} \mu}^{'} {\psi}_{\rm l}^t \widetilde{v}_{\nu}^{\ast} - 
a_{2 \rm L}^{'} v_{{\rm R} \mu \nu}^{'} {\psi}_{\rm L}^t
 + a_{2 \rm L} {\psi}_{\rm L}^t  \widetilde{v}_{\mu \nu}^{\ast} \right] \; . 
\nonumber 
\end{eqnarray}
We obtain the following non-commutative extension of the 
Yukawa interaction (\ref{c12}):
\begin{equation}
{\cal S}_{\rm HFY} = - \int {\md}^4 x \left( \bar{\hat{\chi}}_{\rm R} G 
\star {\hat \Psi}_{1 \rm L}^t \star \widehat{{\Phi}^{\ast}} + 
\bar{\hat{\eta}}_{\rm R} \widetilde{G}  \circ {\hat \Psi}_{2 \rm L}^t \circ
{{\hat \Phi}_2}^{\ast} +  {\rm h.c.} \right) \; . 
\end{equation}

Seiberg--Witten maps constructed with opposite products can be used also in
models based upon tensor product representations. 
The non-commutative extension of the infinitesimal parameter 
${\hat \Omega}_{( \widetilde{\lambda}, {\lambda}_{\rm R}^{'} )}$ using
the opposite Moyal product (\ref{c18}) is (in an appropriate gauge)
\begin{equation}
{\hat \Omega}_{( \widetilde{\lambda}, {\lambda}_{\rm R}^{'} )} = 
\widetilde{\lambda} + 
{\lambda}_{\rm R}^{'} - \frac{1}{4} {\theta}^{\mu \nu} \left\{  
{\partial}_{\mu} \left( \widetilde{\lambda} + {\lambda}_{\rm R}^{'} \right) , 
\widetilde{v}_{\nu} +  v_{{\rm R} \nu}^{'} \right\}  \; \label{c13}.
\end{equation} 
Beside the Seiberg--Witten maps ${\hat \Psi}_{\rm L}$, $\widehat{\varphi
{\chi}_{\rm R}}$  transforming under (\ref{c5}) and using the usual star
product, one  has to introduce a second map ${\hat \Psi}_{\rm L}^{'}$ for
${\psi}_{\rm L}$,  as well as the non-commutative field 
$\widehat{\widetilde{\varphi} {\eta}_{\rm R}}$, transforming under 
(\ref{c13}) as
\begin{equation} \delta \widehat{\varphi {\eta}_{\rm R}} = \imu 
{\hat \Omega}_{( \widetilde{\lambda}, {\lambda}_{\rm R}^{'} )} \circ 
\widehat{\varphi {\eta}_{\rm R}} \; , \qquad \qquad
\delta \hat{\Psi}_{\rm L}^{'} = \imu 
{\hat \Omega}_{( \widetilde{\lambda}, {\lambda}_{\rm R}^{'} )} \circ 
\hat{\Psi}_{\rm L}^{'} \; .
\end{equation}
The complete non-commutative Yukawa interaction is thus
\begin{equation}
{\cal S}_{\rm TPY} = - \int {\md}^4 x \left( \bar{\hat{\Psi}}_{\rm L} 
G^{+} \star \widehat{\varphi {\chi}_{\rm R}} + 
{\bar{\hat{\Psi}}_{\rm L}}^{'} 
{\widetilde{G}}^{+} \circ \widehat{\widetilde{\varphi} {\eta}_{\rm R}} + 
{\rm h.c.} \right) \; . 
\end{equation}

In ref. \cite{Cal} the Seiberg--Witten map of the Higgs field in Yukawa
interaction is different from the map entering the non-commutative extension of
the Higgs action. The gauge fields involved in both maps are different and
have different non-commutative representatives. Moreover, in the minimal 
Non-Commutative Standard Model, the gauge kinetic term is extended to the
non-commutative space-time by means of the Seiberg--Witten  map
\begin{equation}
U_{\rho} \equiv g' \frac{y}{2} B_{\rho} + g \frac{\tau^a}{2} W_{\rho}^a + g_s
\frac{\lambda^A}{2} G_{\rho}^A \; \qquad \mbox{with 
$\; y \equiv \left( \begin{array}{cc}
                 1 &   0  \\ 
                 0 & - 1 
                  \end{array} \right)$} \; ,
\end{equation} 
which is different from all previous non-commutative gauge field
representatives.

In this work we will take a somewhat different point of view. We shall make 
the Seiberg--Witten maps entering Yukawa interaction dynamical, that is we
shall use these maps to perform the extension of all the other terms of the
Standard Model action. Since Yukawa interaction in non-commutative space-time
may involve different Seiberg--Witten maps of equivalent matter
representations, we associate certain weights to their non-commutative
contribution to the extended action. Such a treatment also applies  to the
gauge sector and is similar to the approach advocated in the nmNCSM version. 

The main ingredients to make Seiberg--Witten maps dynamical are non-commutative 
covariant derivatives (for matter fields) and field
strengths (for gauge fields). For instance, the gauge map associated to the
non-commutative tensor field ${\hat \Psi}_{\rm L}$  transforms under
(\ref{c15}) according to  
\begin{equation}
\delta {\hat V}_{\rm L \rho} = {\partial}_{\rho} 
{\hat \Omega}_{( \lambda, {\lambda}_{\rm R} )} + \imu \left[
{\hat \Omega}_{( \lambda, {\lambda}_{\rm R} )} \ster{\star} 
{\hat V}_{\rm L \rho} \right] \; . 
\end{equation}
The dependence of the map ${\hat V}_{\rm L \rho}$ on the commuting gauge
fields can be deduced from the corresponding consistency condition. Up to first
order in  ${\theta}^{\mu \nu}$ and in the gauge defined by (\ref{c5}) one gets
\begin{eqnarray}
{\hat V}_{\rm L \rho} 
& = &  v_{\rho} + v_{\rm R \rho} +  \frac{1}{2} {\theta}^{\mu \nu} 
\left[ - \frac{1}{2} \left\{ v_{\mu} + v_{\rm R \mu} 
, {\partial}_{\nu} \left( v_{\rho} + v_{\rm R \rho} \right) 
+ v_{\nu \rho} + v_{\rm R \nu \rho} \right\} \right. \; \nonumber \\
&   & \mbox{} + \left. D_{\rho} \left(  b_{\rm L} v_{\mu \nu} + 
+ b_{\rm L}^{'} v_{{\rm R} \mu \nu} \right) \right] \;  \label{c11}
\end{eqnarray}  
with $b_{\rm L}$ and $b_{\rm L}^{'}$ being arbitrary real scaling parameters. 

Of course, similar considerations apply to Seiberg--Witten tensor maps
constructed with the opposite  Moyal product.

The covariant derivative of the hybrid maps can be obtained from their
transformation properties. According to (\ref{c3}) the scalar hybrid map 
has the covariant derivative
\begin{equation} 
{\hat {\cal D}}_{\rho} \star {\hat \Phi} 
 = {\partial}_{\rho} {\hat \Phi} -  \imu {\hat V}_{{\rm L} \rho} \star 
{\hat \Phi} + \imu{\hat \Phi} \star {\hat V}_{{\rm R} \rho} \; .
\end{equation}
${\hat V}_{{\rm L} \rho}$ and ${\hat V}_{{\rm R} \rho}$ stand here for the 
Seiberg--Witten maps of the gauge fields $v_{{\rm L} \rho}$ and 
$v_{{\rm R} \rho}$, respectively, and become dynamical together with 
${\hat \Psi}_{\rm L}$ and ${\hat \chi}_{\rm R}$ defined in (\ref{c14}).

%% file: nc4.tex
\section{Models Based on Non-Commutative Tensor Products}
\subsection*{The Tensor Product Model}
The simplest non-commutative version of the Standard Model is
obtained by representing the left chiral leptons and quarks by tensor
Seiberg--Witten maps. The action of this model,  hereafter called the direct
tensor product model, is given by
\begin{eqnarray}
{\cal S}_{\rm PM} 
& = &   \int {\md}^4 x \left\{ {\bar{\hat L}}_{\rm L} \star \imu 
\left( {\hat {\slashed{\cal D}}} \star {\hat L}_{\rm L} \right) + 
{\bar{\hat Q}}_{\rm L} \star \imu 
\left( {\hat {\slashed{\cal D}}} \star {\hat Q}_{\rm L} \right)
\right. \; \nonumber \\
&   & + \mbox{} {\bar{\hat e}}_{\rm R} \star \imu 
\left( {\hat {\slashed{\cal D}}} \star 
{\hat e}_{\rm R} \right) +  {\bar{\hat d}}_{\rm R} \star \imu 
\left( {\hat {\slashed{\cal D}}} \star {\hat d}_{\rm R} \right) 
+ {\bar{\hat u}}_{\rm R} \star \imu 
\left( {\hat {\slashed{\cal D}}} \star {\hat u}_{\rm R} \right)
 \; \nonumber \\
&   & - \mbox{} \left( \bar{\hat{L}}_{\rm L} G_e^{+} 
\star \widehat{\varphi e_{\rm R}} +  \bar{\hat{Q}}_{\rm L} G_d^{+} 
\star \widehat{\varphi d_{\rm R}} + \bar{\hat{Q}}_{\rm L} G_u^{+} 
\star \widehat{\widetilde{\varphi} u_{\rm R}}
 + {\rm h.c.} \right) \; \label{d1} \\    
&   & + \mbox{} \sum_i  w_i \left[ {\left(
{\hat {\cal D}}^{\rho} \star {\hat \Phi}_i \right)}^{+} \star 
\left( {\hat {\cal D}}_{\rho} \star {\hat \Phi}_i \right) + 
\frac{\lambda v^2}{4} 
{\hat \Phi}_i^{+} \star {\hat \Phi}_i - \frac{\lambda}{4} 
{\hat \Phi}_i^{+} \star {\hat \Phi}_i \star {\hat \Phi}_i^{+} \star 
{\hat \Phi}_i \right] \; \nonumber \\
&  & - \mbox{} \left. \frac{1}{2} \sum_{n \neq \varphi} \frac{1}{h_n^2} 
\Tr {\hat V}_{\rho \sigma} ( n ) \star {\hat V}^{\rho \sigma} ( n ) - 
\frac{1}{2 h_{\varphi}^2} \sum_i w_i \Tr {\hat V}_{i \rho \sigma} 
\star {\hat V}_i^{\rho \sigma} \right\} \; . \nonumber  
\end{eqnarray}
where $w_1$ and $w_2$ with $w_1 + w_2 = 1$ are the weights of the maps  
${\hat \Phi}_1$ and  ${\hat \Phi}_2$ of the Higgs field $\varphi$ 
and its charge conjugate $\widetilde{\varphi}$. The summation 
variable $n$ labels the (inequivalent) matter representations of the
Standard Model  gauge group, i.e. $n = L, Q, e, d, u, \varphi$. 
The non-commutative gauge
fields corresponding to the equivalent representations for the Higgs are
denoted by ${\hat V}_1$  and ${\hat V}_2$. 
In order to agree with the normalization of the commutative Standard Model 
we set
\begin{equation}
\frac{1}{h_1^2} = \frac{w_1}{h_{\varphi}^2} \; , \qquad \qquad 
\frac{1}{h_2^2} = \frac{w_2}{h_{\varphi}^2} \; 
\end{equation}
for the weights of their contribution to the kinetic gauge term in the 
action. 
The gauge fields associated to left and right chiral fermions satisfy the
following relations: 
\begin{equation}
v_{\rho} ( L ) = v_{\rho} ( e ) + v_{\rho} ( \varphi ) \; , \quad \qquad 
v_{\rho} ( Q ) = v_{\rho} ( d ) + v_{\rho} ( \varphi ) = 
v_{\rho} ( u ) + v_{\rho} ( \widetilde{\varphi} ) \; .
\end{equation}

The action for a typical matter sector of the tensor product model can be
written in the form
\begin{eqnarray}
{\cal S}_{\rm PM}^{(m)} 
& = &   \int {\md}^4 x \left[ {\bar{\hat \Psi}}_{\rm L} \star \imu 
\left( {\hat {\slashed{\cal D}}} \star {\hat \Psi}_{\rm L} \right) + 
 {\bar{\hat \chi}}_{\rm R} \star \imu 
\left( {\hat {\slashed{\cal D}}} \star 
{\hat \chi}_{\rm R} \right) +  {\bar{\hat \eta}}_{\rm R} \star \imu 
\left( {\hat {\slashed{\cal D}}} \star {\hat \eta}_{\rm R} \right) 
\right. \; \nonumber \\
&   & - \mbox{} \left. \left( \bar{\hat{\Psi}}_{\rm L} G^{+} 
\star \widehat{\varphi \chi_{\rm R}} + 
\bar{\hat{\Psi}}_{\rm L} {\widetilde{G}}^{+} 
\star \widehat{\widetilde{\varphi} \eta_{\rm R}}
 + {\rm h.c.} \right) \right] \;  \label{d4}
\end{eqnarray}
where, for simplicity, we omit the label $m$ of fermionic Seiberg--Witten
maps.  
Since the couplings of the  left chiral fermion to the Higgs and its charge
conjugated are  different, one is expecting two distinct non-commutative maps, 
$\hat{\Psi}_{1 \rm L}$ and $\hat{\Psi}_{2 \rm L}$. According to  (\ref{c10})
they may differ only by terms proportional to the partial field strengths
${v}_{\mu \nu}$, $v_{\rm R \mu \nu}$ and $\widetilde{v}_{\mu \nu}$, 
$v_{\rm R \mu \nu}^{'}$. By
taking $a_{\rm L}^{'} = a_{\rm L}$ one can render both maps equivalent since
they depend only on the sum $v_{\rm L \mu \nu}$ (see (\ref{c1})).  Similarly,
the choice $b_{\rm L}^{'} = b_{\rm L}$ in (\ref{c11}) leads to the 
equivalence of the corresponding gauge maps  ${\hat V}_{1 \rm L \rho}$ and
${\hat V}_{2 \rm L \rho}$. 
A similar treatment for the Seiberg--Witten product maps 
$\widehat{\varphi {\chi}_{\rm R}}$  and  
$\widehat{\widetilde{\varphi} {\eta}_{\rm R}}$ is not possible because the
scaling  parameters $\alpha$ and ${\alpha}_{\rm R}$ in (3.21) are independent
of each other. They can be however related to the scaling parameters of 
${\hat \Phi}_1$ and  ${\hat \chi}_{\rm R}$ (or ${\hat \Phi}_2$ and 
${\hat \eta}_{\rm R}$). To get  the precise relation we may use the equations
of motion in the matter sector, as explained in Sect. 2. Since the scaling
parameters appear at the first order in ${\theta}^{\mu \nu}$, it is sufficient
to consider the equations of motion  of the commutative Standard Model
\begin{eqnarray}
\imu \slashed{D} \chi_{\rm R} 
& = & G \varphi^{+} \psi_{\rm L} \; , \quad \qquad \imu \slashed{D} 
\eta_{\rm R} = \widetilde{G} \widetilde{\varphi} \psi_{\rm L} \; , \quad 
\qquad \imu \slashed{D} \psi_{\rm L}
= G^{+} \varphi \chi_{\rm R} + {\widetilde G}^{+} 
\widetilde{\varphi} \eta_{\rm R} \; , \nonumber \\
D^{\rho} D_{\rho} \varphi  
& = &  \frac{\lambda v^2}{4} \varphi - 
\frac{\lambda}{2} \left( {\varphi}^{+} \varphi \right) \varphi + 
\sum_m \left[ \psi_{\rm L} {\bar \chi}_{\rm R} G - \imu \tau^2 {\left(
\eta_{\rm R}  {\bar \psi}_{\rm L} {\widetilde{G}}^{+} \right)}^t \right] 
\; .\label{d5}  
\end{eqnarray}
We insert now the Seiberg--Witten maps with arbitrary scaling parameters into
$\sum_m {\cal S}_{\rm PM}^{(m)} +  {\cal S}_{\rm PM}^{(H)}$  and compute the 
non-commutative contribution to first order in  ${\theta}^{\mu \nu}$. 
After using (\ref{d5}), the various scaling factors have to be chosen 
as to cancel out,  independently of $w_1$, $w_2$. 

Let us give now the complete list of the maps occurring in (\ref{d1}):     
\begin{eqnarray}
\hat{\Psi}_{\rm L} 
& = &  {\psi}_{\rm L} + \frac{1}{2} {\theta}^{\mu \nu} \left[ 
v_{\rm L \mu} \left( - {\partial}_{\nu} + \frac{\imu}{2} v_{\rm L \nu} \right)
+ a_{\rm L} v_{{\rm L} \mu \nu} \right] {\psi}_{\rm L} \; , \nonumber \\ 
\widehat{\varphi {\chi}_{\rm R}} 
& = & \varphi {\chi}_{\rm R} + \frac{1}{2} {\theta}^{\mu \nu} \left[ 
v_{\rm L \mu} \left( - {\partial}_{\nu} + \frac{\imu}{2} v_{\rm L \nu} \right)
\left( \varphi {\chi}_{\rm R} \right) + \imu D_{\mu}\varphi D_{\nu} 
{\chi}_{\rm R} \right. \; \nonumber \\
&   & \mbox{} + \left. \left( a v_{\mu \nu} + a_{\rm R} 
v_{\rm R \mu \nu}  \right)\varphi {\chi}_{\rm R} \right] \; , \nonumber \\
\widehat{\widetilde{\varphi} {\eta}_{\rm R}} 
& = &  \widetilde{\varphi} {\eta}_{\rm R} + \frac{1}{2} {\theta}^{\mu \nu}
\left[  v_{\rm L \mu} \left( - {\partial}_{\nu} + \frac{\imu}{2} v_{\rm L \nu}
\right) \left( \widetilde{\varphi} {\eta}_{\rm R} \right) + 
\imu D_{\mu}\widetilde{\varphi} D_{\nu} 
{\eta}_{\rm R} \right. \; \nonumber \\
&   & \mbox{} + \left. \left( - a^{\ast} \widetilde{v}_{\mu \nu} + a_{\rm
R}^{'}  v_{\rm R \mu \nu}^{'}  \right)\widetilde{\varphi} {\eta}_{\rm R}
\right] \; , \nonumber  \\
{\hat \chi}_{\rm R} 
& = & \chi_{\rm R} + 
\frac{1}{2} {\theta}^{\mu \nu} 
\left[  v_{{\rm R} \mu} \left( - {\partial}_{\nu} + 
\frac{\imu}{2}  v_{{\rm R} \nu} \right) 
+ a_{\rm R} v_{{\rm R} \mu \nu} \right] \chi_{\rm R} \; , \label{d8}  \\
{\hat \eta}_{\rm R}     
& = & \eta_{\rm R} + 
\frac{1}{2} {\theta}^{\mu \nu} \left[  
v_{{\rm R} \mu}^{'} \left( - {\partial}_{\nu} + \frac{\imu}{2}  
v_{{\rm R} \nu}^{'} \right) +  a_{\rm R}^{'} 
v_{{\rm R} \mu \nu}^{'} \right] \eta_{\rm R} \; , \nonumber \\
{\hat \Phi}_1   
& = & \varphi + \frac{1}{2} {\theta}^{\mu \nu} 
\left[  v_{\mu} \left( - {\partial}_{\nu} + \frac{\imu}{2}  v_{\nu} \right) 
+ a v_{\mu \nu} \right] \varphi \; , \nonumber \\
{\hat \Phi}_2   
& = & \widetilde{\varphi} + \frac{1}{2} {\theta}^{\mu \nu} 
\left[  \widetilde{v}_{\mu} \left( - {\partial}_{\nu} + 
\frac{\imu}{2}  \widetilde{v}_{\nu} \right) 
- a^{\ast} \widetilde{v}_{\mu \nu} \right] \widetilde{\varphi} 
\; , \nonumber \\
{\hat V}_{1 \rho} 
& = & v_{\rho} + \frac{1}{2} {\theta}^{\mu \nu} 
\left( - \frac{1}{2} \left\{ v_{\mu} , {\partial}_{\nu} v_{\rho} +
v_{\nu \rho} \right\} + b v_{\mu \nu} \right) \; , \nonumber \\ 
{\hat V}_{2 \rho} 
& = & \widetilde{v}_{\rho} + \frac{1}{2} {\theta}^{\mu \nu} 
\left( - \frac{1}{2} \left\{ \widetilde{v}_{\mu} , {\partial}_{\nu} 
\widetilde{v}_{\rho} + \widetilde{v}_{\nu \rho} \right\} + b D_{\rho} 
\widetilde{v}_{\mu \nu} \right) \; , \nonumber \\
 {\hat V}_{\rho} ( n ) 
& = &  v_{\rho} ( n ) +  \frac{1}{2} {\theta}^{\mu \nu} 
\left[ - \frac{1}{2} \left\{ v_{\mu} ( n ) , 
{\partial}_{\nu} v_{\rho} ( n ) 
+ v_{\nu \rho} ( n ) \right\} +  b D_{\rho} v_{\mu \nu} ( n )  \right] \; ,
\nonumber   
\end{eqnarray}
for $n = \rm L, \rm R, {\rm R}'$. While each matter sector (fermionic and
scalar) has its own scaling factor ($a_{\rm L}$ and $a$), the parameter $b$
appearing in the Seiberg--Witten maps of the gauge fields is universal. 

Hence the tensor product model is consistent and one can eliminate by 
appropriate field redefinitions the scaling parameters from all
Seiberg--Witten maps. 

The non-commutative contribution of (\ref{d4}) can be written as 
\begin{equation}
\Delta {\cal S}_{\rm PM}^{(m)} = \int {\md}^4 x \Delta {\cal L}_{\rm PM}^{(m)} 
= \int {\md}^4 x \left( \Delta {\cal L}_{\rm L} + \Delta {\cal L}_{\rm R} + 
\Delta {\cal L}_{\rm R}^{'} - \Delta {\cal L}_{\rm Y} - \Delta {\cal L}_{\rm Y}^{'} \right) \;
\end{equation}
where
\begin{eqnarray}
\Delta {\cal L}_{\rm L} & \equiv & - \frac{\imu}{2} \bar{\psi}_{\rm L} 
{\theta}^{\mu \nu \rho}  v_{\rm L \mu \nu} D_{\rho} \psi_{\rm L} \; ,
\nonumber \\ 
\Delta {\cal L}_{\rm R} & \equiv & - \frac{\imu}{2} \bar{\chi}_{\rm R} 
{\theta}^{\mu \nu \rho}  v_{\rm R \mu \nu} D_{\rho} \chi_{\rm R} \; , \quad 
\qquad \Delta {\cal L}_{\rm R}^{'} \equiv - \frac{\imu}{2} \bar{\eta}_{\rm R} 
{\theta}^{\mu \nu \rho}  v_{\rm R \mu \nu}^{'} D_{\rho} \eta_{\rm R}
\; , \nonumber \\
\Delta {\cal L}_{\rm Y} & \equiv & \frac{1}{2} {\theta}^{\mu \nu} 
\bar{\chi}_{\rm R} G \left( \imu D_{\mu} {\varphi}^{+} D_{\nu} - \frac{1}{2} 
v_{\rm R \mu \nu} {\varphi}^{+} \right) \psi_{\rm L} + \rm h . \rm c . \; ,
\label{d9} \\
\Delta {\cal L}_{\rm Y}^{'} & \equiv & \frac{1}{2} {\theta}^{\mu \nu} \bar{\eta}_{\rm R} 
\widetilde{G} \left( \imu D_{\mu} {\widetilde{\varphi}}^{+} D_{\nu} 
- \frac{1}{2} v_{\rm R \mu \nu}^{'} {\widetilde{\varphi}}^{+} 
\right) \psi_{\rm L} + {\rm h. c.} \; , \nonumber 
\end{eqnarray}
with $\; v_{\rm L \mu \nu} = v_{\mu \nu} + v_{\rm R \mu \nu} =
\widetilde{v}_{\mu \nu} + v_{\rm R \mu \nu}^{'} \;$.

In a similar way one obtains the contribution to the Higgs action 
\begin{equation}
\Delta {\cal S}_{\rm PM}^{(H)} = \left( w_1 - w_2 \right)
\int {\md}^4 x \Delta {\cal L}_H  \;
\end{equation}
where
\begin{eqnarray}
\Delta {\cal L}_H 
& \equiv & \frac{1}{2} {\theta}^{\mu \nu} \left( D_{\nu}
{\varphi}^{+} v_{\mu \rho} D^{\rho} \varphi + {\rm h.c.} - \frac{1}{2} 
 D_{\rho}{\varphi}^{+} v_{\mu \nu} D^{\rho} \varphi \right. \; \nonumber \\ 
&   & \mbox{}- \left. \frac{\lambda v^2}{8}{\varphi}^{+} 
v_{\mu \nu} \varphi - \frac{\lambda}{2} {\varphi}^{+} \varphi 
\imu D_{\mu}{\varphi}^{+} D_{\nu} \varphi \right)
\; . \label{d7}
\end{eqnarray}   

Finally, the non-commutative contribution to the gauge kinetic term is given
by  
\begin{equation}
\Delta {\cal S}_{\rm PM}^{(g)} = \int {\md}^4 x \left[ \sum_{n \neq \varphi} 
\frac{1}{h_n^2} \Delta {\cal L}_{n} + \frac{w_1 - w_2}{h_{\varphi}^2}  
\Delta {\cal L}_{\varphi}  \right] \;
\end{equation}
where 
\begin{equation}
\Delta {\cal L}_{n} \equiv \frac{1}{2} {\theta}^{\mu \nu} 
\Tr v^{\rho \sigma}( n ) \left[ - 2 v_{\mu \rho} ( n ) v_{\nu \sigma} ( n ) + 
\frac{1}{2} v_{\rho \sigma} ( n ) v_{\mu \nu} ( n ) \right] \; \label{d10} 
\end{equation}
for $n = L, Q, e, d, u, \varphi$.

We would like to express these corrections in terms of physical fields. Since 
the gauge symmetry of the Standard Model is spontaneously broken to 
$U_{\rm em} ( 1 ) \bigotimes SU_{\rm c} (  3 )$ we shall use the 
covariant derivative
\begin{equation}
{\nabla}_{\rho} \equiv {\partial}_{\rho} + \imu g_{\rm s} G_{\rho} + \imu 
 \mbox{\boldmath $Q$} {\rm e} A_{\rho} \;  
\end{equation}
with $\;G_{\rho} = T^A G_{\rho}^A \;$. 
Here $\mbox{\boldmath $Q$}$ is a charge operator in flavor space 
with the eigenvalues $Q_{\nu} = 0$, $Q_{e} = - 1$, 
$Q_{u} = 2 / 3$ and $Q_{d} = - 1 / 3$. The corresponding field strengths
are $G_{\rho \sigma} \equiv T^A G_{\rho \sigma}^A$ and $F_{\rho \sigma}$. 
 We will also find it convenient to include a 
coupling constant in the definition of the $Z$-field and its field strength: 
\begin{equation}
{\cal Z}_{\rho} \equiv \frac{g}{2 \cos \theta} Z_{\rho} \; , \qquad \qquad 
 {\cal Z}_{\rho \sigma} = {\partial}_{\rho} {\cal Z}_{\sigma} - 
{\partial}_{\sigma} {\cal Z}_{\rho} \; .
\end{equation} 
Another useful abbreviation is
\begin{equation}
W_{\rho \sigma}^{\pm} \equiv {\nabla}_{\rho} W_{\sigma}^{\pm} -
{\nabla}_{\sigma} W_{\rho}^{\pm}  \; 
\end{equation}
where $\; {\nabla}_{\rho} W_{\sigma}^{\pm}
=  {\partial}_{\rho} W_{\sigma}^{\pm} \pm \imu {\rm e} 
A_{\rho} W_{\sigma}^{\pm} \;$.
The non-commutative contribution to the full fermionic sector consists of a
flavor changing  (FC) and a flavor preserving (FP) part
\begin{equation}
\sum_m \Delta {\cal S}^{(m)} = \int {\md}^4 x \left( \Delta {\cal L}^{\rm FC} + 
{\rm h. c.} + \Delta {\cal L}^{\rm FP} \right) \; . \label{d14}
\end{equation}   
This decomposition holds for all models under consideration, so we have
omitted the subscript. Furthermore, we separate the lepton ($l$) from 
the quark ($q$) flavor changing contribution:
\begin{equation}
\Delta {\cal L}^{\rm FC} = \Delta {\cal L}^{\rm l} + \Delta {\cal L}^{\rm q} \; . 
\label{d15} 
\end{equation}
We obtain the following expressions:
\begin{eqnarray}
\Delta {\cal L}_{\rm PM}^{\rm l}
& = & W_{\rho}^{-} \frac{g}{\sqrt{2}}  \bar{e}  
{\theta}^{\mu \nu \rho} \left( \imu \stackrel{\leftarrow}{\nabla}_{\mu}
{\nabla}_{\nu} -  {\cal Z}_{\nu} \stackrel{\leftarrow}{\nabla}_{\mu}  - \cos ( 2
\theta ) {\nabla}_{\nu} {\cal Z}_{\mu} + \frac{\rm e}{2} F_{\mu \nu} \right)
{\nu}_{\rm L}  \; \nonumber \\
&   & + \mbox{} ( H + v ) \frac{g}{\sqrt{2}}  \bar{e} 
\frac{1}{2} {\theta}^{\mu \nu} 
\frac{M_e}{v} \left[ W_{\mu}^{-} \left( \stackrel{\leftarrow}{\nabla}_{\nu} 
- 2 \imu {\cal Z}_{\nu} \right)  \right] 
{\nu}_{\rm L} \; , \label{d11} 
\end{eqnarray}
\begin{eqnarray}
\Delta {\cal L}_{\rm PM}^{\rm q}
& = & W_{\rho}^{-} \frac{g}{\sqrt{2}} \bar{d} V^{+}  
{\theta}^{\mu \nu \rho} \left[ \imu \stackrel{\leftarrow}{\nabla}_{\mu}
{\nabla}_{\nu} +  \left( \frac{4}{3} \sin^2 \theta - 1 \right)
{\cal Z}_{\nu} \stackrel{\leftarrow}{\nabla}_{\mu}  
\right. \; \nonumber \\
&   & - \mbox{} \left. \left( 1 - \frac{2}{3} \sin^2 \theta \right) 
{\nabla}_{\nu} {\cal Z}_{\mu} - g_{\rm s} G_{\mu \nu} -  \frac{\rm e}{6} 
F_{\mu \nu} \right] P_{\rm L} u \; \nonumber \\
&   & + \mbox{} ( H + v ) \frac{g}{\sqrt{2}} \bar{d} 
\frac{1}{2} {\theta}^{\mu \nu} \left\{ 
\frac{M_d}{v} V^{+} P_{\rm L} W_{\mu}^{-}
\left[ \stackrel{\leftarrow}{\nabla}_{\nu} -   
2 \imu \left( 1 - \frac{2}{3} \sin^2 \theta \right) {\cal Z}_{\nu} \right] 
W_{\mu}^{-} \right. \; \nonumber \\
&   & + \mbox{} \left. V^{+} \frac{M_u}{v} P_{\rm R} W_{\mu}^{-}
\left[  {\nabla}_{\nu}  - 2 \imu \left( 1 - \frac{1}{3} 
\sin^2 \theta  \right) {\cal Z}_{\nu} \right] W_{\mu}^{-} 
\right\} u  \; ,
\end{eqnarray}

\begin{eqnarray}
\Delta {\cal L}_{\rm PM}^{\rm FP}
& = & \sum_f \bar{f} \frac{1}{2} {\theta}^{\mu \nu \rho} \left\{ \imu 
\left[ g_{\rm s} G_{\mu \nu} + \mbox{\boldmath $Q$} {\rm e} F_{\mu \nu} + 
2 \left( \mbox{\boldmath $T_3$} P_{\rm L} - \mbox{\boldmath $Q$} \sin^2 
\theta \right) {\cal Z}_{\mu \nu} \right] \right. \; \nonumber \\
&   & \times \mbox{} \left.\left[ {\nabla}_{\rho} + 2 \imu 
\left( \mbox{\boldmath $T_3$} P_{\rm L} - \mbox{\boldmath $Q$} \sin^2 
\theta \right) {\cal Z}_{\rho} \right] - \frac{g^2}{4} \left( W_{\mu}^{-} 
W_{\nu \rho}^{+} + \rm h. \rm c. \right) P_{\rm L} \right\} f \; 
\nonumber \\
&   & + \mbox{} g^2 W_{\mu}^{+}W_{\nu}^{-} \sum_f 
\bar{f} \frac{1}{2} {\theta}^{\mu \nu \rho} \mbox{\boldmath $T_3$}\left\{
\stackrel{\leftarrow}{\nabla}_{\rho} + {\nabla}_{\rho} \right. 
\; \label{d12} \\
&   &  + \mbox{} \left. 4 \imu \left[ 
\mbox{\boldmath $Q$} \sin^2 \theta - \mbox{\boldmath $T_3$} \left( 2 + 
\cos ( 2 \theta ) \right) \right] {\cal Z}_{\rho} \right\} P_{\rm L} f \; 
\nonumber \\
&   & + \mbox{} ( H + v ) \sum_f \bar{f} \frac{1}{2} {\theta}^{\mu \nu} 
\frac{\mbox{\boldmath $M$}}{v} \left[ \imu \stackrel{\leftarrow}{\nabla}_{\mu}
{\nabla}_{\nu} + 2 \left( \mbox{\boldmath $T_3$} P_{\rm L} - 
\mbox{\boldmath $Q$} \sin^2 \theta \right) {\cal Z}_{\mu}  
\stackrel{\leftarrow}{\nabla}_{\nu} \right. \; \nonumber \\
&   & + \mbox{} \left. 2 \left( \mbox{\boldmath $T_3$} P_{\rm R} - 
\mbox{\boldmath $Q$} \sin^2 \theta \right) {\nabla}_{\nu} {\cal Z}_{\mu} -     
g_{\rm s} G_{\mu \nu} - \mbox{\boldmath $Q$} {\rm e} F_{\mu \nu} 
 + \imu g^2 \mbox{\boldmath $T_3$} W_{\mu}^{+} W_{\nu}^{-} \right] f 
\;  \nonumber   
\end{eqnarray}
where the covariant derivative $\stackrel{\leftarrow}{\nabla}$  
is acting on all the factors after the coupling
constant, or after the summation symbol. In (\ref{d12}) we introduced two new
operators diagonal in flavor space,  the  third component of the weak isospin
$\mbox{\boldmath $T_3$}$ with eigenvalues  $T_{3 \nu} = 1 / 2$, $T_{3 e} = - 1
/ 2$, $T_{3 u} = 1 / 2$ and $T_{3 \nu} = - 1 / 2$, and the mass operator
$\mbox{\boldmath $M$}$ with nontrivial  eigenvalues given by 
\begin{equation}
M_e = {\rm diag} \left( m_e, m_{\mu}, m_{\tau} \right) \; , \quad 
M_u = {\rm diag} \left( m_u, m_c, m_t \right) \; , \quad 
M_d = {\rm diag} \left( m_d, m_s, m_b \right) \; .\label{d6} 
\end{equation}

Since the Higgs contribution is proportional to the weight difference, tensor
product models include the case when the Higgs sector does not contribute 
at all, to first order in ${\theta}^{\mu \nu}$. Furthermore, the action of
nmNCSM exhibited in \cite{MPTSW} is obtained by taking $w_1 = 1$ and $w_2 = 0$.
In order to facilitate the comparison we present (\ref{d7}) in terms of
physical fields:
\begin{eqnarray}
\Delta {\cal L}_H 
& = &\frac{1}{2} {\theta}^{\mu \nu}
\left\{ \frac{g^2}{2} W_{\mu}^{+} W^{- \rho} \left[ \imu 
{\partial}_{\nu} H {\partial}_{\rho} H  + ( H + v ) {\cos}^2 \theta 
{\partial}_{\nu} H {\cal Z}_{\rho} \right. \right. \; \nonumber \\*
&   & \mbox{} + \left. \frac{1}{2} ( H + v )^2 \left( {\rm e} F_{\nu \rho} + 
\cos ( 2 \theta ) {\cal Z}_{\nu \rho} + 2 \imu ( 1 + \cos^2 \theta )
{\cal Z}_{\nu}{\cal Z}_{\rho} \right)  \right] \; \nonumber \\
&   & \mbox{} + \frac{g^2}{4} W_{\mu \rho}^{+} \left[ \imu ( H + v ) 
\left( W_{\nu}^{-}
{\partial}^{\rho} H + W^{- \rho} {\partial}_{\nu} H  -
\frac{1}{2} {\delta}_{\nu}^{\rho} W^{-} \cdot {\partial} H \right) \right. 
\; \nonumber \\
&   & \mbox{} + \left. 
( H + v )^2 \left( W_{\nu}^{-} {\cal Z}^{\rho} + W^{- \rho} {\cal Z}_{\nu} 
-  \frac{1}{2} {\delta}_{\nu}^{\rho} W^{-} \cdot {\cal Z} \right) \right] + 
{\rm h. c.}  \; \label{d16} \\ 
&   & \mbox{} + {\cal Z}_{\mu \rho} \left[ {\partial}_{\nu} H {\partial}^{\rho}
H  - \frac{1}{4} {\delta}_{\nu}^{\rho} ( {\partial} H )^2  + 
( H + v )^2 \left( {\cal Z}_{\nu} {\cal Z}^{\rho} - \frac{1}{4} 
{\delta}_{\nu}^{\rho} {\cal Z}^2 \right) 
\right. \; \nonumber \\ 
&   & \mbox{} + \left.  \frac{\lambda}{32} {\delta}_{\nu}^{\rho} 
H^2 ( H + 2 v )^2  \right] + g^2 W^{+} \cdot W^{-} \left[ - ( H + v ) {\cos}^2
\theta {\cal Z}_{\mu} {\partial}_{\nu} H  \right. \; \nonumber \\
&   & \mbox{} + \left. \frac{1}{8} ( H + v )^2 
\left( {\rm e} F_{\mu \nu} + \cos ( 2 
\theta ) {\cal Z}_{\mu \nu} \right) \right]
- \frac{\imu g^2}{4} W_{\mu}^{+} W_{\nu}^{-} 
\left[ ( \partial H )^2 + \frac{\imu}{4} ( H + v )^2 \right.\; \nonumber \\
&   & \mbox{} \times \left.\left. \left( 
( 1 + 4 \cos^2 \theta ) {\cal Z}^2 + 
\frac{g^2}{2} W^{+} W^{-} - \frac{\lambda}{4} H ( H + 2 v ) \right) \right] 
\right\} \; . \nonumber    
\end{eqnarray} 

The non-commutative version of the Standard Model gauge sector has been
thoroughly  discussed in \cite{{Cal},{AJSW},{BDDSTW}} and the results
can be taken over with minor changes (at least for $w_1 \neq w_2$). 

\subsection*{The Twisted Product Model}
The commutative action for the Higgs field is invariant under the
replacement $\varphi \longleftrightarrow \widetilde{\varphi} = \imu \tau^2 
{\varphi}^{\ast}$. By promoting this property to the non-commutative
space-time, the Seiberg--Witten maps of  $\widetilde{\varphi}$ and 
$\varphi$ are related by complex  conjugation and have to be
constructed with opposite Moyal products. 
The left-chiral quark, which couples to both $\varphi$ and
$\widetilde{\varphi}$ will be represented by two Seiberg--Witten tensor maps
${\hat Q}_{\rm L}$ and  ${\hat Q}_{\rm L}^{'}$, complex conjugate each other.
Similarly, the tensor products  $d_{\rm R} \varphi$ and $u_{\rm R}
\widetilde{\varphi}$ will be represented by maps constructed with opposite 
Moyal products.
As a consequence, the non-commutative fields of $u_{\rm R}$, $v_{\rho}  ( u )$
and $v_{\rho} ( u ) + \widetilde{v}_{\rho}$ will be constructed  with
the opposite product, while those of $d_{\rm R}$, $v_{\rho} ( d )$ and
$v_{\rho}  ( u ) + v_{\rho}$ will use the usual star product.

The action of this non-commutative version of the Standard Model, called here
the twisted product model is given by 
\begin{eqnarray}
{\cal S}_{\rm TM} 
& = &   \int {\md}^4 x \left\{ {\bar{\hat L}}_{\rm L} \star \imu 
\left( {\hat {\slashed{\cal D}}} \star {\hat L}_{\rm L} \right) + 
w {\bar{\hat Q}}_{\rm L} \star \imu 
\left( {\hat {\slashed{\cal D}}} \star {\hat Q}_{\rm L} \right) +  
w' {\bar{\hat Q}}_{\rm L}^{'} \circ \imu 
\left( {\hat {\slashed{\cal D}}} \circ {\hat Q}_{\rm L}^{'} \right) 
\right. \; \nonumber \\
&   & + \mbox{} {\bar{\hat e}}_{\rm R} \star \imu 
\left( {\hat {\slashed{\cal D}}} \star 
{\hat e}_{\rm R} \right) +  {\bar{\hat d}}_{\rm R} \star \imu 
\left( {\hat {\slashed{\cal D}}} \star {\hat d}_{\rm R} \right) 
+ {\bar{\hat u}}_{\rm R} \circ \imu 
\left( {\hat {\slashed{\cal D}}} \circ {\hat u}_{\rm R} \right)
 \; \nonumber \\
&   & - \mbox{} \left( \bar{\hat{L}}_{\rm L} G_e^{+} 
\star \widehat{\varphi e_{\rm R}} +  \bar{\hat{Q}}_{\rm L} G_d^{+} 
\star \widehat{\varphi d_{\rm R}} + \bar{\hat{Q}}_{\rm L}^{'} G_u^{+} 
\circ \widehat{\widetilde{\varphi} u_{\rm R}}
 + {\rm h.c.} \right) \; \nonumber \\    
&   & + \mbox{}   {\left(
{\hat {\cal D}}^{\rho} \star {\hat \Phi} \right)}^{+} \star 
\left( {\hat {\cal D}}_{\rho} \star {\hat \Phi} \right) + 
\frac{\lambda v^2}{4} 
{\hat \Phi}^{+} \star {\hat \Phi} - \frac{\lambda}{4} 
{\hat \Phi}^{+} \star {\hat \Phi} \star {\hat \Phi}^{+} \star 
{\hat \Phi}  \; \nonumber \\
&   & - \mbox{}  \frac{1}{2} {\sum_n}^{'}  \frac{1}{h_n^2} 
\Tr {\hat V}_{\rho \sigma} ( n ) \star {\hat V}^{\rho \sigma} ( n ) -
\frac{1}{2 h_u^2} 
\Tr {\hat V}_{\rho \sigma} ( u ) \circ {\hat V}^{\rho \sigma} ( u ) 
\; \nonumber \\
&   & - \mbox{} \left. 
\frac{1}{2 h_Q^2} \Tr \left[ w {\hat V}_{\rho \sigma} ( Q ) 
\star {\hat V}^{\rho \sigma} ( Q ) + w' {\hat V}_{\rho \sigma}^{'} ( Q ) 
\circ {\hat V}^{' \rho \sigma} ( Q ) \right] \right\} \;   
\end{eqnarray}   
where $w$, $w'$ (with $w + w' = 1$) are the weights of the tensor maps for
the left-chiral quark field and the primed sum goes over $n = L, e, d, \varphi$. 
In order to check the consistency of this model we write the action of the 
general fermionic sector in a form similar to (\ref{d4})
\begin{eqnarray}
{\cal S}_{\rm TM}^{(m)} 
& = &   \int {\md}^4 x \left[ w {\bar{\hat \Psi}}_{\rm L} \star \imu 
\left( {\hat {\slashed{\cal D}}} \star {\hat \Psi}_{\rm L} \right) + 
w'{\bar{\hat \Psi}}_{\rm L}^{'} \circ \imu 
\left( {\hat {\slashed{\cal D}}} \circ {\hat \Psi}_{\rm L}^{'} \right) + 
 {\bar{\hat \chi}}_{\rm R} \star \imu 
\left( {\hat {\slashed{\cal D}}} \star 
{\hat \chi}_{\rm R} \right)  
\right. \; \nonumber \\
&   & + \mbox{} \left. {\bar{\hat \eta}}_{\rm R} \circ \imu 
\left( {\hat {\slashed{\cal D}}} \circ {\hat \eta}_{\rm R} \right) -  
\left( \bar{\hat{\Psi}}_{\rm L} G^{+} 
\star \widehat{\varphi \chi_{\rm R}} + 
\bar{\hat{\Psi}}_{\rm L}^{'} {\widetilde{G}}^{+} 
\circ \widehat{\widetilde{\varphi} \eta_{\rm R}}
 + {\rm h.c.} \right) \right] \; . 
\end{eqnarray}  
By going through the same steps as before, one can establish the independent
scaling parameters. The Seiberg--Witten maps using the star product can be read
out from  (\ref{d8}). We record here only the maps constructed with the
opposite Moyal product:
\begin{eqnarray}
\hat{\Psi}_{\rm L}^{'} 
& = &  {\psi}_{\rm L} + \frac{1}{2} {\theta}^{\mu \nu} \left[ -  
v_{\rm L \mu} \left( - {\partial}_{\nu} + \frac{\imu}{2} v_{\rm L \nu} \right)
+ a_{\rm L} v_{{\rm L} \mu \nu} \right] {\psi}_{\rm L} \; , \nonumber \\ 
{\hat \eta}_{\rm R}     
& = & \eta_{\rm R} + 
\frac{1}{2} {\theta}^{\mu \nu} \left[ -   
v_{{\rm R} \mu}^{'} \left( - {\partial}_{\nu} + \frac{\imu}{2}  
v_{{\rm R} \nu}^{'} \right) +  a_{\rm R}^{'} 
v_{{\rm R} \mu \nu}^{'} \right] \eta_{\rm R} \; , \nonumber \\
\widehat{\widetilde{\varphi} {\eta}_{\rm R}} 
& = &  \widetilde{\varphi} {\eta}_{\rm R} + \frac{1}{2} {\theta}^{\mu \nu}
\left[ - v_{\rm L \mu} \left( - {\partial}_{\nu} + \frac{\imu}{2} v_{\rm L \nu}
\right) \left( \widetilde{\varphi} {\eta}_{\rm R} \right) - 
\imu D_{\mu}\widetilde{\varphi} D_{\nu} 
{\eta}_{\rm R} \right. \; \\
&   & \mbox{} + \left. \left( - a^{\ast} \widetilde{v}_{\mu \nu} + 
a_{\rm R}^{'}  v_{\rm R \mu \nu}^{'}  \right)\widetilde{\varphi} {\eta}_{\rm R}
\right] \; , \nonumber \\ 
 {\hat V}_{\rho} ( n ) 
& = &  v_{\rho} ( n ) +  \frac{1}{2} {\theta}^{\mu \nu} 
\left[ \frac{1}{2} \left\{ v_{\mu} ( n ) , 
{\partial}_{\nu} v_{\rho} ( n ) 
+ v_{\nu \rho} ( n ) \right\} +  b D_{\rho} v_{\mu \nu} ( n )  \right] \; 
\nonumber 
\end{eqnarray}
for $n = {\rm L}', {\rm R}' \;$.

With the notations introduced in (\ref{d9}), (\ref{d7}) and (\ref{d10}), the 
non-commutative contributions to the action of the twisted product model take
the following form:
\begin{eqnarray}
\Delta {\cal S}_{\rm TM}^{(m)} 
& = & \int {\md}^4 x \left[ ( w - w' ) \Delta {\cal L}_{\rm L} + \Delta {\cal L}_{\rm R} 
- \Delta {\cal L}_{\rm R}^{'} - \Delta {\cal L}_{\rm Y} + \Delta {\cal L}_{\rm Y}^{'} \right) 
\; , \nonumber \\
\Delta {\cal S}_{\rm TM}^{(H)} 
& = & \int {\md}^4 x \Delta {\cal L}_H  \; \\
\Delta {\cal S}_{\rm TM}^{(g)} 
& = & \int {\md}^4 x \left[ {\sum_n}' 
\frac{1}{h_n^2} \Delta {\cal L}_{n} + \frac{w - w'}{h_Q^2}  
\Delta {\cal L}_Q  - \frac{1}{h_u^2} \Delta {\cal L}_{u} \right] \; . 
\nonumber
\end{eqnarray}

When expressed through physical fields, the whole leptonic contribution
remains unchanged with respect to the direct product model, e.g. 
$\Delta {\cal L}_{\rm TM}^{\rm l} = \Delta {\cal L}_{\rm PM}^{\rm l}$ as given 
by (\ref{d11}). The quark contribution to the flavor changing part is
\begin{eqnarray}
\Delta {\cal L}_{\rm TM}^{\rm q}
& = &  W_{\rho}^{-} \frac{g}{\sqrt{2}} ( w - w' ) \bar{d} V^{+}  
{\theta}^{\mu \nu \rho} \left[ \imu \stackrel{\leftarrow}{\nabla}_{\mu}
{\nabla}_{\nu} +  \left( \frac{4}{3} \sin^2 \theta - 1 \right)
{\cal Z}_{\nu} \stackrel{\leftarrow}{\nabla}_{\mu}  
\right. \; \nonumber \\
&   & - \mbox{} \left. \left( 1 - \frac{2}{3} \sin^2 \theta \right) 
{\nabla}_{\nu} {\cal Z}_{\mu} - g_{\rm s} G_{\mu \nu} -  \frac{\rm e}{6} 
F_{\mu \nu} \right] P_{\rm L} u \; \nonumber \\
&   & + \mbox{} ( H + v ) \frac{g}{\sqrt{2}}  \bar{d} 
\frac{1}{2} {\theta}^{\mu \nu} \left\{ 
\frac{M_d}{v} V^{+} P_{\rm L} W_{\mu}^{-} 
\left[ \stackrel{\leftarrow}{\nabla}_{\nu}  - 
2 \imu \left( 1 - \frac{2}{3} \sin^2 \theta \right) {\cal Z}_{\nu} \right] 
W_{\mu}^{-} \right. \; \nonumber \\
&   & - \mbox{} \left. V^{+} \frac{M_u}{v} P_{\rm R}
\left[ {\nabla}_{\nu}  - 2 \imu \left( 1 - \frac{1}{3} 
\sin^2 \theta  \right) {\cal Z}_{\nu} \right] W_{\mu}^{-} 
\right\} u  \; .
\end{eqnarray}

The contribution of the flavor preserving part to the extended action can be 
also written in compact way as follows:
\begin{eqnarray}
\Delta {\cal L}_{\rm TM}^{\rm FP}
& = & \sum_f \bar{f} {\theta}^{\mu \nu \rho} \mbox{\boldmath $T_3$} 
\left\{ \imu 
\left[ - \left( g_{\rm s} G_{\mu \nu} + \mbox{\boldmath $Q$} {\rm e} F_{\mu \nu} 
  - 2 \mbox{\boldmath $Q$} \sin^2 \theta  
{\cal Z}_{\mu \nu} \right) \left( \gamma_5 + 2 \mbox{\boldmath $w$} P_{\rm L} 
\right) \right. \right. \; \nonumber \\*
&   & + \mbox{} \left.  2 ( 1 - 2 \mbox{\boldmath $w$} ) 
\mbox{\boldmath $T_3$} P_{\rm L} {\cal Z}_{\mu \nu} \right] 
\left[ {\nabla}_{\rho} + 2 \imu 
\left( \mbox{\boldmath $T_3$} P_{\rm L} - \mbox{\boldmath $Q$} \sin^2 
\theta \right) {\cal Z}_{\rho} \right]   \; \nonumber \\
&   & - \mbox{} \left. \frac{g^2}{4} ( 1 - 2 
\mbox{\boldmath $w$} ) \left( W_{\mu}^{-} 
W_{\nu \rho}^{+} + \rm h. \rm c. \right) P_{\rm L} \right\} f 
\; \nonumber \\
&   & + \mbox{} \frac{g^2}{4} W_{\mu}^{+}W_{\nu}^{-} \sum_f 
\bar{f} {\theta}^{\mu \nu \rho} ( 1 - 2 \mbox{\boldmath $w$} )
\left\{
\stackrel{\leftarrow}{\nabla}_{\rho} + {\nabla}_{\rho} \right. 
\;  \label{d13} \\
&   & + \mbox{} \left. 4 \imu \left[ 
\mbox{\boldmath $Q$} \sin^2 \theta - \mbox{\boldmath $T_3$} \left( 2 + 
\cos ( 2 \theta ) \right) \right] {\cal Z}_{\rho} \right\} P_{\rm L} f \; 
\nonumber \\
&   & - \mbox{} ( H + v ) \sum_f \bar{f} \frac{1}{2} {\theta}^{\mu \nu}
2 \mbox{\boldmath $T_3$} 
\frac{\mbox{\boldmath $M$}}{v} \left[ \imu \stackrel{\leftarrow}{\nabla}_{\mu}
{\nabla}_{\nu} + 2 \left( \mbox{\boldmath $T_3$} P_{\rm L} - 
\mbox{\boldmath $Q$} \sin^2 \theta \right) {\cal Z}_{\mu}  
\stackrel{\leftarrow}{\nabla}_{\nu} \right. \; \nonumber \\
&   & + \mbox{} \left. 2 \left( \mbox{\boldmath $T_3$} P_{\rm R} - 
\mbox{\boldmath $Q$} \sin^2 \theta \right) {\nabla}_{\nu} {\cal Z}_{\mu} -     
g_{\rm s} G_{\mu \nu} - \mbox{\boldmath $Q$} {\rm e} F_{\mu \nu} 
 - \imu g^2 \mbox{\boldmath $T_3$} W_{\mu}^{+} W_{\nu}^{-} \right] f 
\; \nonumber  
\end{eqnarray}
where we introduced a weight operator $\mbox{\boldmath $w$}$ diagonal in
flavor space with eigenvalues $w_{\nu} = 0$, $w_e = 1$, $w_u = w'$ 
and $w_d = w$. It is a simple exercise to check that the leptonic parts 
of (\ref{d12}) and (\ref{d13}) coincide. 

A peculiarity of the twisted product model is that left- and right-handed
chiral quarks carry different charges in non-commutative space-time and, as a
consequence, the electromagnetic interactions violate parity.  To see this we 
compute from (\ref{d13}) the electromagnetic interaction to
lowest order  in the coupling constant ${\rm e}$, by assuming the quarks on
their mass-shell. We get
\begin{eqnarray}
&& A_{\rho} \frac{{\rm e}}{6} {\theta}^{\mu \nu} \left\{ \bar d \left[ 
{\delta}_{\mu}^{\rho} \left( 2 w' \stackrel{\leftarrow}{\partial}_{\nu} 
M_d P_{\rm R} + {\rm h.c.} \right) - \imu 
\stackrel{\leftarrow}{\partial}_{\mu} \gamma^{\rho} \left( 1 - 2 w' 
P_{\rm L} \right) {\partial}_{\nu} \right] d \right. \; \nonumber \\
&& + \mbox{} \left. 2  \bar u \left[ 
{\delta}_{\mu}^{\rho} \left( 2 w \stackrel{\leftarrow}{\partial}_{\nu} 
M_u P_{\rm R} + {\rm h.c.} \right) - \imu 
\stackrel{\leftarrow}{\partial}_{\mu} \gamma^{\rho} \left( 1 - 2 w 
P_{\rm L} \right) {\partial}_{\nu} \right] u \right\} \; 
\end{eqnarray}
Similar observations apply to strong interactions.

%% file: nc5.tex
\section{Models Based upon Hybrid Seiberg--Witten Map}
\subsection*{The Hybrid Scalar Model}
In this model the Higgs field is represented by three scalar
Seiberg--Witten maps  ${\hat \Phi}_i$ with $i = e, d, u$, hybrid of the left
and right chiral fermions labelled by $i$. The non-commutative action is
\begin{eqnarray}
{\cal S}_{\rm HS} &=& \int {\md}^4 x \left\{ {\bar{\hat L}}_{\rm L} \star 
\imu \left( \hat{\slashed{\cal D}} \star \hat{L}_{\rm L} \right) + 
{\bar{\hat e}}_{\rm R} \star \imu \left( \hat{\slashed{\cal D}} \star 
{\hat e}_{\rm R} \right) +  {\bar{\hat Q}}_{\rm L} \star \imu 
\left( \hat{\slashed{\cal D}} \star {\hat Q}_{\rm L} \right) 
\right.\; \nonumber \\  
&& \mbox{} + {\bar{\hat d}}_{\rm R} \star \imu \left( 
\hat{\slashed{\cal D}} 
\star {\hat d}_{\rm R} \right) + {\bar{\hat u}}_{\rm R} \star \imu 
\left( \hat{\slashed{\cal D}} \star {\hat u}_{\rm R} \right) - \left( 
{\bar{\hat L}}_{\rm L} G_e^{+} \star {\hat \Phi}_e \star 
{\hat e}_{\rm R} \right. \; \nonumber \\
&&  \mbox{} + \left. {\bar{\hat Q}}_{\rm L} G_d^{+} \star 
{\hat \Phi}_d \star {\hat d}_{\rm R} + 
{\bar{\hat Q}}_{\rm L} G_u^{+} \star {\hat \Phi}_u \star 
{\hat u}_{\rm R} + {\rm h}.{\rm c}. \right) \; \nonumber \\
 && \mbox{} + \sum_i  w_i \Tr \left[ 
\left( {\hat{\cal D}}^{\rho} \star {\hat \Phi}_i \right)^{+}
\star \left( {\hat{\cal D}}_{\rho} \star {\hat \Phi}_i \right) + 
\frac{\lambda v^2}{4} {\hat \Phi}_i^{+} \star {\hat \Phi}_i - 
\frac{\lambda}{4} 
{\hat \Phi}_i^{+} \star {\hat \Phi}_i \star {\hat \Phi}_i^{+} \star 
{\hat \Phi}_i  \right]  \; \nonumber \\
&& \mbox{} - \left. \frac{1}{2} \sum_n \frac{1}{h_n^2} 
\Tr {\hat V}_{\rho \sigma} ( n ) \star {\hat V}^{\rho \sigma} ( n ) 
\right\} \;  
\end{eqnarray}
where $w_i$, with $i = e, d, u$ are the weights of the Higgs representatives, 
$\sum_i w_i = 1$. Since hybrid maps do not have their own gauge
Seiberg--Witten maps, we {\em assume} that the kinetic gauge term receives
contributions only from $n = L, Q, e, d, u$. 

The consistency check of the model leads to a single (complex) parameter for 
all the three hybrid maps. Denoting by ${\hat \Phi}$, ${\hat \Phi}'$ the
representatives of the Higgs field $\varphi$ and its charge conjugate 
$\widetilde{\varphi}$, one finds (compare to (\ref{c16}))
\begin{eqnarray}
{\hat \Phi}     & = & \phi + \frac{1}{2} {\theta}^{\mu \nu} 
\left[  v_{{\rm L} \mu} \left( - {\partial}_{\nu} + \frac{\imu}{2}  
v_{{\rm L} \nu} \right)\phi  +  \phi \left( 
\stackrel{\leftarrow}{\partial}_{\mu} + \frac{\imu}{2} 
v_{{\rm R} \mu} \right) v_{{\rm R} \nu} \right. \; \nonumber \\
                &   & \mbox{} - \left.     
\imu v_{{\rm L} \mu} \phi v_{{\rm R} \nu} + a \left( 
v_{{\rm L} \mu \nu} \phi
 - \phi  v_{{\rm R} \mu \nu} \right) \right]  \; , \nonumber \\
{\hat \Phi}^{'} & = & \widetilde{\phi} + \frac{1}{2} {\theta}^{\mu \nu} 
\left[  v_{{\rm L} \mu} \left( - {\partial}_{\nu} + \frac{\imu}{2}  
v_{{\rm L} \nu} \right)\widetilde{\phi}  +  \widetilde{\phi} \left( 
\stackrel{\leftarrow}{\partial}_{\mu} + \frac{\imu}{2} 
v_{{\rm R} \mu}^{'} \right) v_{{\rm R} \nu}^{'} \right. \; \nonumber \\  
                 &   & \mbox{} - \left.    
\imu v_{{\rm L} \mu} \widetilde{\phi} v_{{\rm R} \nu}^{'} + a \left( 
v_{{\rm L} \mu \nu} \widetilde{\phi}
 - \widetilde{\phi}  v_{{\rm R} \mu \nu}^{'} \right) \right]  \; 
\end{eqnarray}
where $\phi = \varphi \otimes {\rm I}_3$ and 
$\widetilde{\phi} = \widetilde{\varphi} \otimes {\rm I}_3$. Field
redefinitions can be now performed owing to the 
identities
\begin{equation}
v_{{\rm L} \mu \nu} \phi = \phi  v_{{\rm R} \mu \nu} + v_{\mu \nu} \varphi 
\otimes {\rm I}_3 \; , \quad \qquad 
v_{{\rm L} \mu \nu} \widetilde{\phi} = \widetilde{\phi}  
v_{{\rm R} \mu \nu}^{'} + \widetilde{v}_{\mu \nu} \widetilde{\varphi} 
\otimes {\rm I}_3 \; .
\end{equation}

The leptonic and quark sectors of the hybrid scalar and of the tensor product
models coincide. The Higgs sector contributes to the extended action with the 
following expression:
\begin{equation}
\Delta {\cal S}_{\rm HS}^{(H)} = \int {\md}^4 x \left[ \left( w_e + w_d - w_u 
\right) \Delta {\cal L}_H  + \left( \frac{2 w_u - w_d}{3} - w_e \right) 
{\cal E}_{\rm HS} \right] \; \label{e5}
\end{equation}
where $\Delta {\cal L}_H$ is given by (\ref{d7}) and
\begin{equation}
{\cal E}_{\rm HS} \equiv \frac{1}{2} {\theta}^{\mu \nu} \left\{ 2 D^{\rho}
{\varphi}^{+} D_{\mu} \varphi + {\rm h.c.} +  \delta_{\mu}^{\rho}
\left[ - D {\varphi}^{+} \cdot D \varphi + \frac{\lambda}{4}
{\varphi}^{+} \varphi 
\left( {\varphi}^{+} \varphi - v^2 \right) \right] \right\} g' B_{\nu \rho} 
 \; . \label{e6}
\end{equation}
The first term in (\ref{e5}) has the same form as Higgs contribution of the
tensor product model if one takes, for instance, $w_1 = w_e + w_d$ and $w_2 =
w_u$, but other combinations are possible. If $3w_d + 5 w_e \neq 2$, the
hybrid  scalar model leads to electromagnetic interactions of neutral particles
($Z$-bosons, Higgs mesons). In terms of physical fields the contribution
(\ref{d6}) takes the form
\begin{eqnarray}
{\cal E}_{\rm HS} 
& = & \frac{1}{2} {\theta}^{\mu \nu} \left\{ 2 
{\partial}_{\mu} H {\partial}^{\rho} H  +  ( H + v )^2 \left( 
\frac{g^2}{2} W_{\mu}^{+} W^{- \rho} + {\rm h.c.} + 2    
{\cal Z}_{\mu} {\cal Z}^{\rho} \right) \right. \; \nonumber \\
&   & - \mbox{} \left. \frac{1}{2} {\delta}_{\mu}^{\rho} \left[ (\partial H )^2 
+ ( H + v )^2 \left( \frac{g^2}{2} W^{+} W^{-} + {\cal Z}^2 \right) - 
\frac{\lambda}{8}  
H^2 ( H + 2 v )^2  \right] \right\} \; \nonumber \\
&   & \times \mbox{} \left( {\rm e} F_{\nu \rho} - 2 \sin^2
\theta  {\cal Z}_{\nu \rho} \right) \; . \label{e9} 
\end{eqnarray}

Electromagnetic interactions of neutral particles occur whenever hybrid maps
become dynamical. In contrast to non-commutative covariant derivatives of
tensor fields, the covariant derivative of a hybrid map destroys the relation
(\ref{c1}) between gauge fields associated to left and right chiral fermions, 
already at first order in ${\theta}^{\mu \nu}$.

\subsection*{The Hybrid Fermion Model}
The hybrid fermion model is based upon extending (\ref{c12}) to
non-commutative space-time. This is achieved by constructing fermionic
Seiberg--Witten maps transforming left and right, as right fermionic singlets
and Higgs doublets, respectively. We decide to use a single Seiberg--Witten
map for the Higgs doublet, although a version with two independent Higgs maps 
${\hat \Phi}_1$, ${\hat \Phi}_2$ of $\varphi$ and $\widetilde{\varphi}$ would
be also conceivable. With the choice described in sect 3. the maps
representing the Higgs fields are 
\begin{eqnarray}
\widehat{{\Phi}^{\ast}} 
& = & {\varphi}^{\ast} + 
\frac{1}{2} {\theta}^{\mu \nu}  \left[  v_{\mu}^{\ast} \left( 
{\partial}_{\nu} + \frac{\imu}{2} v_{\nu}^{\ast} \right)  + a^{\ast} 
v_{\mu \nu}^{\ast} \right]  {\varphi}^{\ast} \; , 
\label{e1} \\ 
{{\hat \Phi}_{2}}^{\ast} 
& = & {\widetilde{\varphi}}^{\ast} + \frac{1}{2} 
{\theta}^{\mu \nu} \left[ - {{\widetilde{v}}_{\mu}}^{\ast} 
\left( - {\partial}_{\nu} + \frac{\imu}{2}{{\widetilde{v}}_{\nu}}^{\ast}
\right)  -  a {{\widetilde{v}}_{\mu \nu}}^{\ast} \right] 
{\widetilde{\varphi}}^{\ast} \; .  \label{e2}
\end{eqnarray} 
Since (\ref{e2}) is obtained from ${\hat \Phi}_{2}$ by complex conjugation, it
is constructed with the opposite Moyal product.

The action of the hybrid fermion model is given by
\begin{eqnarray}
{\cal S}_{\rm HF} 
& = &   \int {\md}^4 x \left\{ {\bar{\hat L}}_{\rm L}^t \star \imu 
\left( {\hat {\slashed{\cal D}}} \star {\hat L}_{\rm L}^t \right) + 
w {\bar{\hat Q}}_{\rm L}^t \star \imu 
\left( {\hat {\slashed{\cal D}}} \star {\hat Q}_{\rm L}^t \right) +  
w' {\bar{\hat Q}}_{\rm L}^{' t} \circ \imu 
\left( {\hat {\slashed{\cal D}}} \circ {\hat Q}_{\rm L}^{' t} \right) 
\right. \; \nonumber \\
&   & + \mbox{} {\bar{\hat e}}_{\rm R} \star \imu 
\left( {\hat {\slashed{\cal D}}} \star 
{\hat e}_{\rm R} \right) +  {\bar{\hat d}}_{\rm R} \star \imu 
\left( {\hat {\slashed{\cal D}}} \star {\hat d}_{\rm R} \right) 
+ {\bar{\hat u}}_{\rm R} \circ \imu 
\left( {\hat {\slashed{\cal D}}} \circ {\hat u}_{\rm R} \right)
 \; \nonumber \\
&   & - \mbox{} \left( \bar{\hat{e}}_{\rm R} G_e \star {\hat L}_{\rm L}^t  
\star \widehat{{\Phi}^{\ast}} + 
\bar{\hat{d}}_{\rm R} G_d \star {\hat Q}_{\rm L}^t  
\star \widehat{{\Phi}^{\ast}} + 
\bar{\hat{u}}_{\rm R} G_u \circ {\hat Q}_{\rm L}^{' t}  
\circ {{\hat \Phi}_2}^{\ast} 
 + {\rm h.c.} \right) \; \nonumber \\    
&   & + \mbox{} {\left(
{\hat {\cal D}}^{\rho} \star {\hat \Phi} \right)}^{+} \star 
\left( {\hat {\cal D}}_{\rho} \star {\hat \Phi} \right) + 
\frac{\lambda v^2}{4} 
{\hat \Phi}^{+} \star {\hat \Phi} - \frac{\lambda}{4} 
{\hat \Phi}^{+} \star {\hat \Phi} \star {\hat \Phi}^{+} \star 
{\hat \Phi}  \; \nonumber \\
&   & - \mbox{} \left. \frac{1}{2} {\sum_n}^{'}  \frac{1}{h_n^2} 
\Tr {\hat V}_{\rho \sigma} ( n ) \star {\hat V}^{\rho \sigma} ( n ) -
\frac{1}{2 h_u^2} 
\Tr {\hat V}_{\rho \sigma} ( u ) \circ {\hat V}^{\rho \sigma} ( u ) 
\right\} \; .  
\end{eqnarray} 
Here ${\hat L}_{\rm L}^t$, ${\hat Q}_{\rm L}^t$, ${\hat L}_{\rm L}^{' t}$ 
are the hybrid Seiberg--Witten maps required by the gauge invariance of Yukawa
couplings and the superscript $t$ means transposition in electroweak
isospin-hypercharge  space. The quark maps are opposite to each other and occur
with weights $w$,  $w'$, obeying $w + w' = 1$. Again we do not consider those
contributions to the gauge sector which are associated to hybrid maps. Hence 
the primed sum goes over $n = e, d, \varphi$.

The consistency of the model can be examined in the same way as in 
previous models. As a consequence, the scaling parameters of the hybrid maps 
${\hat \Psi}_{1 {\rm L}}^t$, ${\hat \Psi}_{2 {\rm L}}^t$ introduced in
(\ref{c6}) are restricted to a single one
\begin{equation}
a_{1 {\rm L}}^{'} = - a_{1 {\rm L}} = - a_{2 {\rm L}} =   
a_{2 {\rm L}}^{'} \equiv a_{\rm L} \; . \label{d2}
\end{equation} 

After rescaling the fields, one can compute the non-commutative contribution of
the various  sectors of the Standard Model in terms of physical fields. Since
both kind of maps (constructed with direct and opposite Moyal products) occur
in the hybrid fermionic model, we expect to get electromagnetic and strong
interactions which violate parity. This is apparent from 
the flavor preserving part contribution
\begin{eqnarray} 
\Delta {\cal L}_{\rm HF}^{\rm FP}
& = & \sum_f \bar{f} {\theta}^{\mu \nu \rho} \mbox{\boldmath $T_3$} 
\left\{ \imu 
\left[ - \left( g_{\rm s} G_{\mu \nu} + \mbox{\boldmath $Q$} {\rm e} F_{\mu \nu} 
  - 2 \mbox{\boldmath $Q$} \sin^2 \theta  
{\cal Z}_{\mu \nu} \right) \left( \gamma_5 + 2 \mbox{\boldmath $w$} P_{\rm L} 
\right) \right. \right. \; \nonumber \\
&   & + \mbox{} \left.  4 ( \mbox{\boldmath $w$} - 1 ) 
\mbox{\boldmath $T_3$} P_{\rm L} {\rm e} F_{\mu \nu} + 2 \left( 1 + 2 
( \mbox{\boldmath $w$} - 1 ) \cos ( 2 \theta ) \right)
\mbox{\boldmath $T_3$} P_{\rm L} {\cal Z}_{\mu \nu} \right] 
\; \nonumber \\
&   & \mbox{} \times \left. \left[ {\nabla}_{\rho} + 2 \imu 
\left( \mbox{\boldmath $T_3$} P_{\rm L} - \mbox{\boldmath $Q$} \sin^2 
\theta \right) {\cal Z}_{\rho} \right] + \frac{g^2}{4} ( 1 - 2 
\mbox{\boldmath $w$} ) \left( W_{\mu}^{-} 
W_{\nu \rho}^{+} + \rm h. \rm c. \right) P_{\rm L} \right\} f 
\; \nonumber \\
&   & - \mbox{} \frac{g^2}{4} W_{\mu}^{+}W_{\nu}^{-} \sum_f 
\bar{f} {\theta}^{\mu \nu \rho} ( 1 - 2 \mbox{\boldmath $w$} )
\left\{
\stackrel{\leftarrow}{\nabla}_{\rho} + {\nabla}_{\rho} \right. 
\; \nonumber \\  
&   & + \mbox{} \left. 4 \imu \left[ 
\mbox{\boldmath $Q$} \sin^2 \theta - \mbox{\boldmath $T_3$} \left( 2 + 
\cos ( 2 \theta ) \right) \right] {\cal Z}_{\rho} \right\} P_{\rm L} f \; 
\nonumber \\
&   & + \mbox{} ( H + v ) \sum_f \bar{f} \frac{1}{2} {\theta}^{\mu \nu}
\frac{\mbox{\boldmath $M$}}{v} \left\{ 2 \mbox{\boldmath $T_3$} 
\left[ \imu \stackrel{\leftarrow}{\nabla}_{\mu}
{\nabla}_{\nu} + 2 \left( \mbox{\boldmath $T_3$} P_{\rm L} - 
\mbox{\boldmath $Q$} \sin^2 \theta \right)   
\stackrel{\leftarrow}{\nabla}_{\nu} {\cal Z}_{\mu} \right. \right. 
\; \nonumber \\ 
&   & + \mbox{} \left. \left. 2 \left( \mbox{\boldmath $T_3$} P_{\rm R} - 
\mbox{\boldmath $Q$} \sin^2 \theta \right) {\cal Z}_{\mu} {\nabla}_{\nu} 
\right] - \frac{1}{2} \left( {\cal Z}_{\mu \nu} +     \imu g^2
W_{\mu}^{+}W_{\nu}^{-} \right) \right\} f \; .  
\end{eqnarray} 
Even the lepton sector is changed in comparison with previous models. 
The electromagnetic interaction of the leptons is given by
\begin{equation}
\frac{1}{2} {\theta}^{\mu \nu} {\rm e} F_{\nu \rho} \left\{ \frac{1}{2} \bar e 
\left[ {\delta}_{\mu}^{\rho}  \left( \imu \slashed{\nabla} - M_e \right) -
2 \imu \gamma^{\rho}{\nabla}_{\mu} \right]  e  +  {\bar \nu}_{\rm L} \left(
 {\delta}_{\mu}^{\rho} \imu \slashed{\partial} - 2 \imu \gamma^{\rho} 
{\partial}_{\mu} \right) \nu_{\rm L} \right\} \; . \label{e7} 
\end{equation} 
The last term of (\ref{e7}) 
describes the photon-neutrino interaction considered in ref. \cite{STWR}. 
Clearly, the  hybrid map of left-chiral leptons 
${\hat L}_{\rm L}^t$ is responsible for such a coupling.

For completeness we give in the following the contributions of the hybrid
fermion model to the flavor changing part of the Standard Model:
\begin{eqnarray}
\Delta {\cal L}_{\rm HF}^{\rm l}
& = & - W_{\rho}^{-} \frac{g}{\sqrt{2}}  \bar{e}  
{\theta}^{\mu \nu \rho} \left( \imu \stackrel{\leftarrow}{\nabla}_{\mu}
{\nabla}_{\nu} - \stackrel{\leftarrow}{\nabla}_{\mu} {\cal Z}_{\nu} - \cos ( 2
\theta ) {\cal Z}_{\mu} {\nabla}_{\nu} - \frac{\rm e}{2} F_{\mu \nu} + 
\sin^2 \theta {\cal Z}_{\mu \nu} \right) {\nu}_{\rm L}  \; \nonumber \\
&   & - \mbox{} ( H + v ) \frac{g}{\sqrt{2}} \bar{e} 
\frac{1}{2} {\theta}^{\mu \nu} 
\frac{M_e}{v} W_{\mu}^{-} \left( \stackrel{\leftarrow}{\nabla}_{\nu} 
 - 2 \imu {\cal Z}_{\nu} \right)   
{\nu}_{\rm L} \;  \\ 
\Delta {\cal L}_{\rm HF}^{\rm q}
& = &  W_{\rho}^{-} \frac{g}{\sqrt{2}} \bar{d} V^{+}  
{\theta}^{\mu \nu \rho} \left\{ - ( w - w' )\left[ \imu 
\stackrel{\leftarrow}{\nabla}_{\mu}
{\nabla}_{\nu} +  \left( \frac{4}{3} \sin^2 \theta - 1 \right)
\stackrel{\leftarrow}{\nabla}_{\mu} {\cal Z}_{\nu} 
\right. \right. \; \nonumber \\
&   & - \mbox{} \left. \left. \left( 1 - \frac{2}{3} \sin^2 \theta \right) 
{\cal Z}_{\mu} {\nabla}_{\nu} \right] +   
\frac{\rm e}{2} F_{\mu \nu} - \sin^2 \theta 
{\cal Z}_{\mu \nu} \right\} P_{\rm L} u \; \nonumber \\
&   & - \mbox{} ( H + v ) \frac{g}{\sqrt{2}} \bar{d} 
\frac{1}{2} {\theta}^{\mu \nu} \left\{ 
\frac{M_d}{v} V^{+} P_{\rm L} W_{\mu}^{-}
\left[ \stackrel{\leftarrow}{\nabla}_{\nu} -  
2 \imu \left( 1 - \frac{2}{3} \sin^2 \theta \right) {\cal Z}_{\nu} \right] 
 \right. \; \nonumber \\
&   & - \mbox{} \left. V^{+} \frac{M_u}{v} P_{\rm R} 
\left[ {\nabla}_{\nu}  - 
2 \imu \left( 1 - \frac{1}{3} \sin^2 \theta  \right) {\cal Z}_{\nu} \right] 
W_{\mu}^{-} \right\} u  \; . 
\end{eqnarray}
 
Finally, the contribution of the model to the gauge sector is
\begin{equation}
\Delta {\cal S}_{\rm HF}^{(g)} = \int {\md}^4 x \left[  
\frac{1}{h_e^2} \Delta {\cal L}_{e} + \frac{1}{h_d^2} \Delta {\cal L}_{d} 
- \frac{1}{h_u^2} \Delta {\cal L}_{u} + \frac{1}{h_{\varphi}^2}  
\Delta {\cal L}_{\varphi}  \right] \; . \label{e8}
\end{equation}
Notice that (\ref{e8}) contains only four weights $1 /  h_n^2$ restricted by
the  normalization conditions (\ref{c17}). 

We expect qualitatively similar conclusions for the alternatively hybrid
fermionic model with two independent non-commutative Higgs fields mentioned at
the beginning of this subsection. In particular, the photon-neutrino
interaction from (\ref{e7}) is the same.

%% file: nc6.tex
\section{Effective Low Energy Four-Fermi Interaction}
In this section we will concentrate on the low energy aspects of the Standard
Model in non-commutative space-time. As in the commutative Standard Model,
energies below the masses of the weak bosons and Higgs (but larger than 
${\Lambda}_{\rm QCD}$) are considered low. Effective actions based upon
Standard Model in non-commutative space-time were first considered in
\cite{T}.  

Here we obtain the effective action by integrating out the weak
massive boson fields  $W_{\rho}^{\pm}$, $Z_{\rho}$ and $H$,
while electromagnetic and strong interactions remain unintegrated. The
corresponding free currents are 
\begin{equation} 
J_{\rho}^{\pm} = \frac{g}{\sqrt{2}} C_{\rho}^{\pm} \; , \quad \qquad 
J_{\rho}^0 = \frac{g}{\cos \theta} C_{\rho}^0 \; , \quad \qquad 
J_H =\frac{1}{v} C_H \; 
\end{equation}
where 
\begin{eqnarray}
& C_{\rho}^{-} \equiv {\bar e} \gamma_{\rho} \nu_{\rm L} + 
{\bar d} V^{+} \gamma_{\rho} 
P_{\rm L} u \; , \qquad \; & C_{\rho}^0 \equiv \sum_f {\bar f} \gamma_{\rho}
\left( \mbox{\boldmath $T_3$} P_{\rm R} - \mbox{\boldmath $Q$} \sin^2 
\theta \right) f \; , \nonumber \\
& C_{\rho}^{+} \equiv {\bar \nu}_{\rm L} \gamma_{\rho} e + {\bar u} V
\gamma_{\rho}  P_{\rm L} d \; , \qquad \; & C_H \equiv \sum_f {\bar f} 
\mbox{\boldmath $M$} f \; .
\end{eqnarray}
Due to the factor $1 / v$, only the top quark contribution to the Higgs 
free current $C_H$ is of comparable order of magnitude as the other free 
currents. Moreover, the contributions of the Higgs path integral to the
effective action are suppressed by a factor $1 / m_H^2$ and will be ignored
in the following.   

The full weak currents are obtained by varying
the extended action with respect to the fields  $W_{\rho}^{\pm}$, $Z_{\rho}$
and $H$.  The part of the Lagrangian due to non-commutative space-time receives
contributions from the  usual three sectors
\begin{equation}
\Delta {\cal L} = \Delta {\cal L}^{\rm matter} + \Delta {\cal L}^{\rm Higgs} + 
\Delta {\cal L}^{\rm gauge} \; .
\end{equation} 
Here $\Delta {\cal L}^{\rm matter} = \Delta {\cal L}^{\rm l} + \Delta {\cal L}^{\rm q} + 
\rm h. \rm c. + \Delta {\cal L}^{\rm FP}$, as given by (\ref{d14}) and
(\ref{d15}), $\Delta {\cal L}^{\rm Higgs}$ is a certain linear combination of 
(\ref{d16}) and (\ref{e9}) and $\Delta {\cal L}^{\rm gauge}$ represents the
contribution of the triple gauge boson interactions ( see \cite{{BDDSTW},
{MPTSW}})  relevant at low energies. 

The construction of the effective action starts by computing the variation
of  $\Delta {\cal L}$ with respect to the massive bosonic field. Then, both
fields and their variations are replaced by the convolution of free field
propagators and the corresponding free currents. The effective action is
obtained as a  power series in inverse masses squared, but 
only the coefficients of lowest powers can be expressed in terms of Standard
Model parameters. All the other coefficients depend in general on the
regularization details.   

It has been shown in Sect. 2 that the contribution of the fermions to the
non-com\-mu\-ta\-tive  effective action does not depend on the regularization
details, because involving the totally antisymmetric tensor ${\theta}^{\mu \nu
\rho}$. This  property continues to be valid for the Standard Model extensions
presented here.  On the other hand, $\Delta {\cal L}^{\rm Higgs}$ and $\Delta {\cal
L}^{\rm gauge}$  contain triple and quadruple boson interactions which are changed
by renormalization. Moreover, such couplings in $\Delta {\cal L}^{\rm Higgs}$ have
positive mass dimension and can contribute to order $1 / m^4$ in the effective
action. Hence we will explicitly integrate only
terms at most quadratically in the massive boson fields, thereby restricting
our considerations to  effective four-fermionic interactions. 

The variation of $\Delta {\cal L}^{\rm matter}$ is then given by     
\begin{equation}
\delta \Delta {\cal L}_{\rm X}^{\rm matter} 
= \delta W_{\rho}^{-} \frac{g}{\sqrt{2}}  \left[ 
{\left({\cal C}_{\rm X}^l\right)}_{+}^{\rho} + 
{\left({\cal C}_{\rm X}^q\right)}_{+}^{\rho} \right] + {\rm h.c.} +  
\delta Z_{\rho} \frac{g}{\cos \theta}  
{\left({\cal C}_{\rm X}^{\rm FP}\right)}_0^{\rho}  \; , \label{f2}  
\end{equation}
provided that the coefficient of $\delta H$ is neglected. In (\ref{f2})
the subscript ${\rm X}$ stands for any of the four models discussed in 
previous  sections. The coefficients of the field variations carry 
superscripts denoting the lepton ($l$), or quark ($q$) contributions to the
flavor changing part, or of both, to the flavor preserving part (FP) for
the non-commutative fermionic matter sector. The coefficients relevant to the
effective four-fermionic interactions are given by
\begin{equation}
{\left({\cal C}_{\rm X}^l\right)}_{+}^{\rho} 
 = \bar{e} \left[  
{\theta}^{\mu \nu \rho} \left( \imu \stackrel{\leftarrow}{\nabla}_{\mu}
{\nabla}_{\nu} + \frac{\rm e}{2} F_{\mu \nu} \right) -  
\frac{1}{2} {\theta}^{\rho \mu} 
M_e {\nabla}_{\mu} \right] {\nu}_{\rm L} \; , \label{f4} 
\end{equation}

\begin{equation}
{\left({\cal C}_{\rm HF}^l\right)}_{+}^{\rho} 
 =  \bar{e} \left[  
{\theta}^{\mu \nu \rho} \left( - \imu \stackrel{\leftarrow}{\nabla}_{\mu}
{\nabla}_{\nu} + \frac{\rm e}{2} F_{\mu \nu} \right) + 
\frac{1}{2} {\theta}^{\rho \mu} 
M_e {\nabla}_{\mu} \right] {\nu}_{\rm L} \; ,
\end{equation}

\begin{eqnarray}
{\left({\cal C}_{\rm X}^q\right)}_{+}^{\rho}
& = &  \bar{d} \left[ V^{+}  
{\theta}^{\mu \nu \rho} \left( \imu \stackrel{\leftarrow}{\nabla}_{\mu}
{\nabla}_{\nu}  - g_{\rm s} G_{\mu \nu} -  \frac{\rm e}{6} 
F_{\mu \nu} \right) P_{\rm L}   \right.\; \nonumber \\
&   & - \mbox{} \left. 
\frac{1}{2} {\theta}^{\rho \mu} \left( M_d V^{+} P_{\rm L} 
{\nabla}_{\mu}  +  V^{+} M_u P_{\rm R} 
\stackrel{\leftarrow}{\nabla}_{\mu} \right) \right]  u , \; 
\end{eqnarray}

\begin{eqnarray}
{\left({\cal C}_{\rm TM}^q\right)}_{+}^{\rho}
& = &  \bar{d} \left[ ( w - w' ) V^{+}  
{\theta}^{\mu \nu \rho} \left( \imu \stackrel{\leftarrow}{\nabla}_{\mu}
{\nabla}_{\nu}  - g_{\rm s} G_{\mu \nu} -  \frac{\rm e}{6} 
F_{\mu \nu} \right) P_{\rm L}   \right.\;  \nonumber \\
&   & - \mbox{} \left. 
\frac{1}{2} {\theta}^{\rho \mu} \left( M_d V^{+} P_{\rm L} 
{\nabla}_{\mu}  -  V^{+} M_u P_{\rm R} 
\stackrel{\leftarrow}{\nabla}_{\mu} \right) \right]  u , \; 
\end{eqnarray}

\begin{eqnarray}
{\left({\cal C}_{\rm HF}^q\right)}_{+}^{\rho}
& = &  \bar{d} \left\{ {\theta}^{\mu \nu \rho} V^{+}  
 \left[ - ( w - w' ) \imu \stackrel{\leftarrow}{\nabla}_{\mu}
{\nabla}_{\nu}  + \frac{\rm e}{2} 
F_{\mu \nu}  \right] P_{\rm L}  \right.\; \nonumber \\
&   & + \mbox{} \left. 
\frac{1}{2} {\theta}^{\rho \mu} \left( M_d V^{+} P_{\rm L} 
{\nabla}_{\mu}  -  V^{+} M_u P_{\rm R} 
\stackrel{\leftarrow}{\nabla}_{\mu} \right) \right\}  u  \; , 
\end{eqnarray}

\begin{eqnarray}
{\left({\cal C}_{\rm X}^{\rm FP}\right)}_0^{\rho}
& = &  \sum_f \bar{f} \left\{  
{\theta}^{\mu \nu \rho} \left( \mbox{\boldmath $T_3$} P_{\rm L} - 
\mbox{\boldmath $Q$} \sin^2 \theta \right)
\left( \imu \stackrel{\leftarrow}{\nabla}_{\mu}
{\nabla}_{\nu}  - g_{\rm s} G_{\mu \nu} -  \mbox{\boldmath $Q$}{\rm e} 
F_{\mu \nu} \right) \right. \;  \\
&   & \mbox{} - \left. \frac{1}{2} {\theta}^{\rho \mu} \mbox{\boldmath $M$}
\left[ \left( \mbox{\boldmath $T_3$} P_{\rm L} - \mbox{\boldmath $Q$} \sin^2 
\theta \right) \nabla_{\mu} +
\left( \mbox{\boldmath $T_3$} P_{\rm R} - \mbox{\boldmath $Q$} \sin^2 
\theta \right) \stackrel{\leftarrow}{\nabla}_{\mu} 
\right] \right\} f \; , \nonumber
\end{eqnarray}

\begin{eqnarray}
{\left({\cal C}_{\rm TM}^{\rm FP}\right)}_0^{\rho}
& = &  2 \sum_f \bar{f} \mbox{\boldmath $T_3$} \left\{ - 
{\theta}^{\mu \nu \rho}  \left( \gamma_5 + 2 \mbox{\boldmath $w$} P_{\rm L}
\right) \left( \mbox{\boldmath $T_3$} P_{\rm L} - \mbox{\boldmath $Q$} \sin^2 
\theta \right) \right. \; \nonumber \\
&   & \mbox{} \times 
\left( \imu \stackrel{\leftarrow}{\nabla}_{\mu}
{\nabla}_{\nu}  - g_{\rm s} G_{\mu \nu} -  \mbox{\boldmath $Q$}{\rm e} 
F_{\mu \nu} \right) \;   \\
&   & \mbox{} + \left. \frac{1}{2} {\theta}^{\rho \mu} \mbox{\boldmath $M$}
\left[ \left( \mbox{\boldmath $T_3$} P_{\rm L} - \mbox{\boldmath $Q$} \sin^2 
\theta \right) \nabla_{\mu} +
\left( \mbox{\boldmath $T_3$} P_{\rm R} - \mbox{\boldmath $Q$} \sin^2 
\theta \right) \stackrel{\leftarrow}{\nabla}_{\mu} 
\right] \right\} f \; , \nonumber
\end{eqnarray}

\begin{eqnarray}
{\left({\cal C}_{\rm HF}^{\rm FP}\right)}_0^{\rho}
& = &  2 \sum_f \bar{f} \mbox{\boldmath $T_3$}
\left\{ {\theta}^{\mu \nu \rho} 
\left[ \left( \left( 1 + 2 ( \mbox{\boldmath $w$} - 1 ) \cos ( 2 \theta ) 
\right) \mbox{\boldmath $T_3$} P_{\rm L} +  2 \sin^2 \theta 
\left( \gamma_5 + 2 \mbox{\boldmath $w$} P_{\rm L} \right) 
 \mbox{\boldmath $Q$} \right) \right.
 \right.  \; \nonumber \\
&   & \mbox{} \times 
\left( \imu \stackrel{\leftarrow}{\nabla}_{\mu} {\nabla}_{\nu} 
- g_{\rm s} G_{\mu \nu} -  \left( \mbox{\boldmath $Q$} - \frac{1}{\sin^2 \theta}
\mbox{\boldmath $T_3$} P_{\rm L} \right) {\rm e} F_{\mu \nu} \right)
\;  \\
&   & \mbox + \left. \left( 1 + ( \mbox{\boldmath $w$} - 1 ) 
\cos^2 \theta \right) \left( \mbox{\boldmath $T_3$} g_{\rm s} G_{\mu \nu}
- \frac{1}{4 \sin^2 \theta} {\rm e} F_{\mu \nu} \right) P_{\rm L} 
\right] \;  \nonumber \\
&   & \mbox{} - \left. \frac{1}{2} {\theta}^{\rho \mu} 
\mbox{\boldmath $M$}
\left[ \left( \mbox{\boldmath $T_3$} P_{\rm L} + \mbox{\boldmath $Q$} \sin^2 
\theta \right) \nabla_{\mu} +
\left( \mbox{\boldmath $T_3$} P_{\rm R} + \mbox{\boldmath $Q$} \sin^2 
\theta \right) \stackrel{\leftarrow}{\nabla}_{\mu} 
\right] \right\} f \; .  \nonumber
\end{eqnarray}
Except for (\ref{f4}) which is valid for ${\rm X} = {\rm PM}, {\rm TM}$ and 
${\rm HS}$, the subscript ${\rm X}$ refers to both direct tensor product 
(${\rm PM}$) and hybrid scalar (${\rm HS}$) models.  

Let us discuss now the contribution of the Higgs sector to the effective
action. A look at (\ref{d16}) shows that it contains triple and quadruple
boson interactions proportional to $v^2$, i.e to mass squared.  
The only contribution quadratic in the massive vector bosons is  
coming from the electromagnetic interaction 
\begin{equation}
{\rm e} m_W^2 \frac{1}{2} {\theta}^{\mu \nu} F_{\nu \rho} \left( 
W_{\mu}^{+} W^{- \rho} + {\rm h. c.} - \frac{1}{2} {\delta}_{\mu}^{\rho} 
W^{+} W^{-} \right) \; . \label{f5} 
\end{equation}
In the hybrid scalar model there is, as follows from (\ref{e5}) and (\ref{e9})
an additional electromagnetic interaction contributing to the effective 
action  
\begin{eqnarray}
&& {\rm e} \frac{1}{2} {\theta}^{\mu \nu} F_{\nu \rho} \left\{ 
2 \partial_{\mu} H \partial^{\rho} H - \frac{1}{2} {\delta}_{\mu}^{\rho} 
\left[ ( \partial H )^2 - \frac{m_H^2}{2} H^2 \right] \right. \; \nonumber \\ 
&& \mbox{} + \left. m_W^2 \left[ 2 \left( W_{\mu}^{+} W^{- \rho} + {\rm h. c.}
+  \frac{1}{\cos^2 \theta} Z_{\mu} Z^{\rho} \right) - {\delta}_{\mu}^{\rho} 
\left( W^{+} W^{-} + \frac{1}{2 \cos^2 \theta} Z^2 \right) \right] \right\} \; 
. \label{f3}
\end{eqnarray}
The last two terms in (\ref{f3}) give a contribution of order 
$1 / m^4$ in the effective action and can be neglected.  

For the same reason one can neglect the contribution of
$\Delta {\cal L}^{\rm gauge}$.   

According to (\ref{g2}) the effective action is obtained from (\ref{f2}) and 
from $\Delta {\cal L}^{\rm Higgs}$ by means of the following replacements:
\begin{equation}
\delta W_{\rho}^{\pm},  W_{\rho}^{\pm} \Rightarrow - \frac{g}{m_W^2 \sqrt{2}} 
C_{\rho}^{\pm} \; , \quad \qquad
\delta Z_{\rho}, Z_{\rho} \Rightarrow - \frac{g}{m_Z^2 \cos \theta} 
C_{\rho}^0 \;  \; . 
\end{equation}
The effective Hamiltonian is given by
\begin{eqnarray}
\Delta {\cal H}_{\rm X}^{\rm eff} 
& = & \frac{4 G_F}{\sqrt{2}} \left\{  
C_{\rho}^{-} \left[ {\left({\cal C}_{\rm X}^l\right)}_{+}^{\rho} + 
{\left({\cal C}_{\rm X}^q\right)}_{+}^{\rho} \right] + {\rm h.c.} + 
2 C_{\rho}^0 {\left({\cal C}_{\rm X}^{\rm FP} \right)}_0^{\rho} \right.
\; \nonumber \\ 
&   & \mbox{} - \frac{1}{2} {\theta}^{\mu \nu} {\rm e} F_{\nu \rho}  
\left[ \left( {\kappa}_{\rm X} + 2 {\mu}_{\rm X} \right) 
\left( C_{\mu}^{+} C^{- \rho} + {\rm h.c.}  - \delta_{\mu}^{\rho} 
\frac{{\kappa}_{\rm X}}{2} C^{+} \cdot C^{-} \right) \right. \; \nonumber \\ 
&   &  \mbox{} + \left. \left. {\mu}_{\rm X} \left( 4 C_{\mu}^0 C^{0 \rho} 
- {\delta}_{\mu}^{\rho} ( C^0)^2 \right) \right] \right\} \; \label{f6}
\end{eqnarray}
where $G_F / {\sqrt{2}} \equiv g^2 / ( 8 m_W^2 )$ is the Fermi weak constant 
and
\begin{eqnarray}
&& {\kappa}_{\rm PM} = w_1 - w_2 \; , \qquad 
{\kappa}_{\rm TM} = {\kappa}_{\rm HF} = 1 \; , \qquad
{\kappa}_{\rm HS} = w_e + w_d -  w_u = 1 - 2 w_u \; , \nonumber \\
&& {\mu}_{\rm PM} = {\mu}_{\rm TM} = {\mu}_{\rm HF} = 0 \; , \qquad 
 {\mu}_{\rm HS} = \frac{2 w_u - w_d}{3} - w_e \; .
\end{eqnarray}
Eq. (\ref{f6}) summarizes the non-commutative contribution 
of the various models to the processes involving four fermions. 
Since $U_{\rm em} \bigotimes SU_{\rm c}$ gauge symmetry is explicitly
preserved, the formula describes also processes with one or more photons or /
and gluons. In computing the corresponding transition probability amplitudes
one must consider, besides the point interaction (\ref{f6}) tree diagrams in
which one of the fermionic lines is off-shell.  

%% file: ncc.tex
\section{Conclusions}

The implementation of Yukawa interactions in 
non-commutative space-time provided several options for extending the Standard
Model, all of them being  based upon the gauge group 
$U_{\rm Y} ( 1 )\bigotimes SU_{\rm L} ( 2 )\bigotimes SU_{\rm c} ( 3 )$.
Each extension was obtained by assigning certain 
Seiberg--Witten maps to the fields entering the commutative Yukawa
interaction. In tensor product models, one of the maps was transforming  
as the tensor product of the other two. Alternatively, the map associated to
the central factor in the Yukawa product transformed left and right, as the
maps corresponding to left- and right-handed fields, respectively. In this way
we obtained hybrid non-commutative  models. A further proliferation of models
appeared  because the couplings to Higgs and
to its charge conjugate field were independent of each other. The complete 
non-commutative extension could be achieved with the same star product,
or through opposite (complex conjugate) Moyal products. In case that
inequivalent Seiberg--Witten maps were associated to the same commutative
matter representation, we assumed a weighted contribution to the
non-commutative action. 

In particular, we derived, at lowest order in ${\theta}^{\mu \nu}$,
the non-minimal Non-Commutative Standard Model within the class of
tensor products models. Also, by making hybrid Seiberg--Witten maps
dynamical, we found that the corresponding non-commutative models always 
predict an electromagnetic coupling of neutral particles, like $Z$-boson,
Higgs meson or neutrino.

We evaluated also 
low energy effective actions for the non-commutative models, by
integrating out massive bosonic degrees of freedom. 
If pure fermionic matter coupled to massive vector bosons of mass 
$m$, we derived an effective action for four- and six-fermion processes, 
valid up to order $1 / m^4$ included. On
the other hand, due to the Higgs mechanism, 
non-commutative extensions of the Standard Model contain, in general,  
triple and quadruple boson couplings of positive mass
dimension.  Such couplings being renormalized were discarded,
reducing  the  validity of the effective actions to order $1 / m^2$.
We obtained formulas for the low energy effective interaction Hamiltonian 
of four fermions and a number of photons and gluons limited only by 
the $U_{\rm em} ( 1 )\bigotimes SU_{\rm c} ( 3 )$ gauge symmetry.

\vspace{3cm}

{\large \bf Aknowledgement}~~ This work has been supported by the Federal
German Ministry for Research BMBF under project 05HT4PSA/6. I thank Thorsten
Feldmann and  Thomas Mannel for interesting and fruitful discussions and
suggestions.

%% file: nc7.tex
\section {Appendix}

We shall give here a formula for the effective action obtained by integrating
out massive bosonic fields coupled to currents. Our starting point is the
action 
\begin{equation}
\frac{1}{2}\xi_i K_{i j} \xi_j + u_i \xi_i  - V ( \xi ) \; .\label{g1}
\end{equation}
In (\ref{g1}) we use a highly compact notation in which space time coordinates 
and field indices are included in Latin indices $i, j, k, \ldots$, summation
occurs whenever an index is repeated. The basic fields (upon one will
integrate) are denoted by $\xi_i$ and the (free) currents by $u_i$, the
symmetric matrix $K_{i j}$ in the kinetic term is nonsingular, its inverse 
being the free propagator. We also assume $V ( \xi )$ to be polynomial of some
degree $N$ in the field $\xi$ and derivatives
\begin{equation}
V ( \xi ) = \sum_{\alpha = 1}^N \frac{1}{\alpha} C_{i_1 \ldots i_{\alpha}} 
\xi_{i_1} \cdots \xi_{i_{\alpha}} \; . \label{g3}
\end{equation} 

The effective action ${\cal S}^{\rm eff} ( u )$ is defined to be the part of
the path integral
\begin{equation}
\frac{1}{\imu} \ln \int \left( \prod_k {\md} \xi_k \right) \exp \left\{ \imu
\left[  \frac{1}{2}\xi_i K_{i j} \xi_j + u_i \xi_i  - V ( \xi ) \right]
\right\} \; \end{equation}
linear in the coefficients $C_{i_1 \ldots i_{\alpha}}$. 

One obtains 
\begin{equation}
{\cal S}^{\rm eff} ( u ) = - \frac{1}{2} u_i {\bar u}_i + 2 {\bar u}_i 
J_i ( 0 ) - \int_0^1 {\md} a \, {\bar u}_i J_i ( a {\bar u} ) - 
{\cal P}_{N - 2} ( \bar{u} )\; \label {g2} 
\end{equation} 
where ${\bar u}_i \equiv K_{i j}^{-1} u_j$ and $J_i ( \xi )$ is the full 
current defined by $\delta V = J_i ( \xi ) \delta \xi_i$. In terms of the 
coefficients introduced in (\ref{g3}) the full current has the following 
expansion:
\begin{equation}
J_i ( \xi ) = C_i + C_{i j} \xi_j + C_{i j k} \xi_j \xi_k + \cdots \; . 
\end{equation}
Finally, ${\cal P}_{N - 2} ( \bar{u} )$ is a certain polynomial of degree $N -
2$ in  $\bar{u}$. For the purpose of this paper the case $N = 6$, considered 
below, is sufficient. We have
\begin{eqnarray}
{\cal P}_4 ( \bar{u} )
& = &  C_{i j k} \frac{K_{i j}^{-1}}{\imu} {\bar u}_k + \frac{3}{4}  
C_{i j k l} \left(  \frac{K_{i j}^{-1}}{\imu} \frac{K_{k l}^{-1}}{\imu} 
+ 2 \frac{K_{i j}^{-1}}{\imu} {\bar u}_k {\bar u}_l \right) 
\; \nonumber \\ 
&   & \mbox{} + C_{i j k l m} \left( 3 \frac{K_{i j}^{-1}}{\imu} 
\frac{K_{k l}^{-1}}{\imu} {\bar u}_m      
+ 2 \frac{K_{i j}^{-1}}{\imu} {\bar u}_k {\bar u}_l {\bar u}_m \right) 
\;  \\ 
&   & \mbox{} +\frac{5}{2} C_{i j k l m n} \left(  
\frac{K_{i j}^{-1}}{\imu} \frac{K_{k l}^{-1}}{\imu} \frac{K_{m n}^{-1}}{\imu}
+ 3 \frac{K_{i j}^{-1}}{\imu} \frac{K_{k l}^{-1}}{\imu} {\bar u}_m {\bar u}_n
+  \frac{K_{i j}^{-1}}{\imu} {\bar u}_k {\bar u}_l {\bar u}_m 
{\bar u}_n \right)  \; . \nonumber \label{g4}
\end{eqnarray}

In a local field theory the factors $K_{i j}^{-1}$ in (\ref{g4}) represent
free propagators at zero distance and are divergent quantities. Without
entering the regularization details one can estimate the 
importance of the various terms in ${\cal P}_4 ( \bar{u} )$ by simple
dimensional analysis. 

Let us apply (\ref{g2}) to get the effective action in non-commutative
space-time for the simple theory described in Sect. 2, where the free currents
$u_i$ are given by  $\bar{\psi} {\gamma}^{\rho} t_a \psi$. Due to the
antisymmetry of  ${\theta}^{\mu \nu \rho}$, the matter sector represented by
(\ref{b7}) does not yield any contribution to the corresponding polynomial. 
Contributions are however expected from the kinetic term for non-commutative 
vector  fields. Notice that each factor ${\bar u}_i$, or $K_{i j}^{-1}$
introduces an inverse power of mass squared. Referring to (\ref{b6}) one 
easily finds that the contribution of (\ref{g4}) to the effective four-fermion 
interaction is given by
\begin{equation}
\frac{3}{2} \left( C_{i j k l} \frac{K_{i j}^{-1}}{\imu} {\bar u}_k {\bar u}_l 
+ 5 C_{i j k l m n} \frac{K_{i j}^{-1}}{\imu} \frac{K_{k l}^{-1}}{\imu} 
{\bar u}_m {\bar u}_n \right) \; . \label{g5}
\end{equation}
Since both $C_{i j k l}$ and $C_{i j k l m n}$ do not introduce factors of
order $m$, the contribution of (\ref{g5}) is of order $1 / m^6$.